\documentclass[aps,prx,amsmath,amssymb,amsfonts,lengthcheck,twocolumn]{revtex4}

\usepackage{epsf}
\usepackage{amsfonts}
\usepackage{amssymb}
\usepackage{graphicx}
\usepackage{color}
\usepackage{amsmath}
\usepackage[bookmarksnumbered, bookmarks, breaklinks, linktocpage]{hyperref}

\newcommand{\Env}{\ensuremath{\mathcal{E}}}

\newcommand{\Tr}        {\mathrm{Tr}}
\newcommand{\bra}[1]    {\langle #1|}
\newcommand{\ket}[1]    {| #1 \rangle}
\newcommand{\bk}[2]     {\langle #1 | #2 \rangle}
\newcommand{\kb}[2]     {| #1 \rangle \! \langle #2 |}
\newcommand{\cH}        {{\mathcal H}}
\newcommand{\cS}        {{\mathcal S}}
\newcommand{\cA}        {{\mathcal A}}
\newcommand{\cE}        {{\mathcal E}}

\newcommand\cF{{\mathcal F}}
\newcommand\hocom[1]{}

\newcommand{\ba}{\begin{eqnarray}}
\newcommand{\ea}{\end{eqnarray}}
\newcommand{\bmath}{\begin{mathletters}}
\newcommand{\emath}{\end{mathletters}}
\newcommand{\ban}{\begin{eqnarray*}}
\newcommand{\ean}{\end{eqnarray*}}

\usepackage{tikz}
\usetikzlibrary{arrows,shapes}

\begin{document}

\tikzstyle{every picture}+=[remember picture]

\title{Quantum Theory of the Classical: Quantum Jumps, Born's Rule, and \\ 
Objective Classical Reality via Quantum Darwinism}

\author{Wojciech Hubert Zurek}

\address{Theory Division, MS B213, LANL
    Los Alamos, NM, 87545, U.S.A.}


\begin{abstract} Emergence of the classical world from the quantum substrate of our Universe is a long-standing conundrum. 
I describe three insights into the transition from quantum to classical that are based on the recognition of the role of the environment. 
I begin with derivation of preferred sets of states that help define what exists - our everyday classical reality. They emerge as a result of breaking of the unitary symmetry of the Hilbert
space which happens when the unitarity of quantum evolutions encounters nonlinearities inherent in the process of amplification -- of replicating information.
 This derivation is accomplished without the usual tools of decoherence, and accounts for the appearance of quantum jumps and emergence of preferred {\it pointer states}
consistent with those obtained via environment-induced superselection, or {\it einselection}. Pointer states obtained this way determine what can happen -- define 
events -- without appealing to Born's rule for probabilities. Therefore, $p_k=|\psi_k|^2$ can be now deduced from the entanglement-assisted invariance, or {\it envariance} 
-- a symmetry of entangled quantum states. With probabilities at hand one also gains new insights into foundations of quantum statistical physics. Moreover, one can now
analyze information flows responsible for decoherence. These information flows explain how perception of objective classical reality arises from the quantum substrate: Effective amplification they represent accounts for the objective existence of the einselected states of macroscopic quantum systems through the redundancy of pointer state records in their
environment -- through {\it quantum Darwinism}. 
\end{abstract}
\maketitle

\section{Introduction and Preview}

This essay is not a comprehensive review. It is nevertheless a brief review of a several interrelated developments that can be collectively described as ``Quantum Theory of Classical Reality''.
Its predecessor 
was intended to be a brief  annotated guide to some of the (then) recent results. 
Present lecture notes evolved away from the annotated guide to literature in the direction of a review in two ways: they are less complete as a guide (as quite a few relevant papers have appeared since 2008, and while I mention some of them, I am sure there are significant omissions). On the other hand, I went beyond the annotated guide canon, and I review some of the key advances in more depth. Still, this is no substitute for the original papers, or for a fully fledged review. 

Two mini-reviews in {\it Nature Physics} \cite{ZurekNatPhys} and {\it Physics Today} \cite{ZurekPT} are also available. A more detailed review is in Ref. \cite{Z07a}. It is by now somewhat out-of-date, as several relevant results were obtained since 2007 when it was written. Moreover, a book that will cover this same ground as the present lectures, as well as theory of decoherence and other related subjects is (slowly) being written \cite{Zurekbook}. Nevertheless, it is hoped that readers may appreciate, in the interim,  an update as well as the more informal presentation style of this overview, including the ``frequently asked questions" in Section VI.

The  ``Relative State Interpretation'' set out 50 years ago by Hugh
Everett III \cite{25,26} is a convenient starting point for our discussion. Within its context, one can reevaluate basic axioms of quantum theory (as extracted, for example, from
Dirac, \cite{23}). Everettian view of the Universe is a good way to motivate exploring the effect of the environment
on the state of the system. (Of course, a complementary motivation based on a non-dogmatic
reading of Bohr \cite{11} is also possible.)

The basic idea we shall pursue here is to accept a relative state explanation of the  ``collapse of the
wavepacket'' by recognizing, with Everett, that observers perceive the state of the ``rest of the Universe''
{\it relative} to their own state, or -- to be more precise -- relative to the state of their records.
This allows quantum theory to be universally valid. (This does {\it not} mean that one has to accept
a ``many worlds'' ontology; see Ref. \cite{Z07a} for discussion.)

Much of the heat in various debates on the foundations of quantum theory seems to be generated by the expectation that a {\it single} idea should provide a complete solution.
When this does not happen -- when there is progress, but there are still unresolved issues -- the possibility that an idea
responsible for this progress may be a step in the right direction -- but that more than one idea, one step, is needed -- is often dismissed.
As we shall see, developing quantum theory of our classical everyday reality requires the solution of {\it several} problems and calls
for several ideas. In order to avoid circularities, they need to be introduced in the right order. 

\subsection{Preferred Pointer States from Einselection}

Everett explains perception of the collapse. However, his relative state approach raises three questions
absent in Bohr's Copenhagen Interpretation \cite{11} that relied on the independent existence
of an {\it ab initio} classical domain. 
Thus, in a completely quantum Universe one is forced to seek
sets of preferred, effectively classical but ultimately quantum states that can define what exists -- branches of the universal
state vector -- and that allow observers keep reliable records. Without such
{\bf preferred basis} relative states are just ``too relative'', and the relative state approach suffers from {\it basis ambiguity} \cite{69}.

Decoherence selects preferred {\it pointer states} \cite{69,70, 71}, so this issue was in fact resolved some time ago.  The principal consequence of the environment-induced decoherence is that, in open quantum systems -- systems interacting with their environments -- only certain quantum states retain stability in spite of the immersion of the system in the environment: Superpositions are unstable, and quickly decay into mixtures of the einselected, stable pointer states \cite{ZurekNatPhys, ZurekPT, Z07a, Zurekbook, 69, Zeh, 70, 71, 45, 36, 72, 73, 75, 51,52}. This is einselection -- a nickname for {\it e}nvironment - {\it in}duced super{\it selection}. Thus, while the significance of the environment in suppressing quantum behavior was pointed out by Dieter Zeh already in 1970 \cite{67}, he role of the einselection in the emergence of these preferred pointer states in the transition from quantum to classical became fully appreciated only since 1981 \cite{ZehDark}.

\subsection{Born's rule from Envariance}

Einselection can account for preferred sets of states, and, hence, for Everettian ``branches''. But this
is achieved at a very high price -- the usual practice of decoherence
is based on averaging (as it involves reduced density matrices defined by a partial trace). This means that
one is using Born's rule to relate amplitudes to probabilities. But, as emphasized by Everett, Born's rule
should not be postulated in an approach that is based on purely unitary quantum dynamics. The assumption of the
universal validity of quantum theory raises the issue of {\bf the origin of Born's rule},
$p_k = |\psi_k|^2$, which -- following the original conjecture \cite{12} -- is simply postulated in textbook discussions.

Here we shall see that Born's rule can be derived from entanglement - assisted invariance, or envariance -- from the symmetry of entangled quantum states. Envariance is a purely quantum symmetry, as it is critically dependent on the telltale quantum feature -- entanglement. Envariance sheds new light on the origin of probabilities relevant for the everyday world we live in, e.g. for statistical physics and thermodynamics. 
Moreover, fundamental derivation of objective probabilities allows one to discuss information flows in our quantum Universe, and, hence, understand how perception of classical reality emerges from quantum substrate.

\subsection{Classical Reality via Quantum Darwinism}

Even preferred quantum states defined by einselection are still ultimately quantum. Therefore, they
cannot be found out by initially ignorant observers through direct measurement without getting
disrupted (reprepared). Yet, states of macroscopic systems in our everyday world seem to exist objectively
-- they can be found out by anyone without getting disrupted. This ability to find out an unknown
state is in fact an operational definition of  ``objective existence". So, if we are to explain the
emergence of everyday objective classical reality, we need to identify {\bf quantum origin of objective existence}.

We shall do that by dramatically upgrading the role of the environment: in decoherence theory the
environment is the collection of degrees of freedom where quantum coherence (and, hence,
phase information) is lost. However, in ``real life'' the role of the environment is in effect that of a witness (see e.g. ~\cite{Z00,75}) to the state of the
system of interest, and a communication channel through which the information
reaches us, the observers. This mechanism for the emergence of the classical objective reality is the subject of the theory of quantum Darwinism.

\section{Quantum Postulates and Relative States}

We start from a well-defined solid ground -- the
list of quantum postulates that are explicit in Dirac \cite{11}, and at least implicit in most quantum textbooks.

The first two deal with the mathematics of quantum theory:

(i) {\it The state of a quantum system is represented by a vector in its Hilbert space} $\cH_{\cS}$.

(ii) {\it Evolutions are unitary (e.g., generated by the Schr\"odinger equation).}

These two postulates provide an essentially complete summary of the mathematical structure of quantum physics.
They are often \cite{21,22} supplemented by a composition postulate: 

(o) {\it States of composite quantum systems are represented by a vector in the tensor product of the Hilbert spaces of its components.}

Physicists sometimes differ in assessing how much of postulate (o) follows from (i).
We shall not be distracted by this issue, and move on to where the real problems are.  Readers
can follow their personal taste in supplementing (i) and (ii) with whatever portion of (o) they deem
necessary. It is nevertheless useful to list (o) explicitly to emphasize the role of the tensor structure it posits: it is crucial for entanglement, quantum phenomenon we will depend on.

Using (o), (i) and (ii), suitable Hamiltonians, etc., one can calculate. Yet, such quantum calculations are only a mathematical exercise 
-- without additional postulates one can predict nothing of experimental consequence from their results. What is so far missing is physics -- a way to establish correspondence 
between abstract state vectors in $\cH_{\cS}$ and laboratory experiments 
(and/or everyday experience) is needed to relate quantum mathematics to our world. 

Establishing this correspondence starts with the next postulate:

(iii) {\it Immediate repetition of a measurement yields the same outcome.}

Immediate repeatability is an idealization (it is hard to devise such non-demolition measurements, but it
can be done). Yet postulate (iii) is uncontroversial. The notion of a ``state'' is based on predictability, and the most rudimentary prediction is 
a confirmation that the state is what it is known to be. This key ingredient of quantum physics goes beyond the mathematics of postulates (o)-(ii). It enters through the repeatability postulate (iii). Moreover, a classical equivalent of 
(iii) is taken for granted (unknown classical state can be discovered without getting disrupted), so repeatability does not clash with our classical intuition.

Postulate (iii) is the last uncontroversial postulate on the textbook list. This collection comprises our {\it quantum core} postulates -- our {\it credo}, the foundation of the quantum theory of the classical. 

In contrast to classical physics (where unknown states can be found out by an initially ignorant observer)
the very next quantum axiom limits the predictive attributes of the state compared to what they were in the classical domain:

(iv) {\it Measurement outcomes are limited to an orthonormal set of states (eigenstates of the measured observable). In any given run of a measurement an outcome is just one such state.}

This {\it collapse postulate} is controversial. To begin with, in a completely quantum Universe
it is inconsistent with the first two postulates: Starting from a general pure state $\ket {\psi_{\cS}}$
of the system (postulate (i)), and an initial state $\ket {A_0}$ of the apparatus $\cA$, and assuming unitary evolution (postulate (ii)) one is led to a superposition of outcomes:
$$\ket {\psi_{\cS}} \ket {A_0} =  \left(\sum_k a_k \ket {s_k}\right) \ket {A_0}
\Rightarrow
\sum_k a_k \ket {s_k} \ket {A_k}
\ , \eqno(1)$$
which is in contradiction with,
at least, a literal interpretation of the ``collapse'' anticipated by axiom (iv). This conclusion follows for an apparatus that works as intended in tests (i.e., $\ket {s_k} \ket {A_0} \Rightarrow \ket {s_k} \ket {A_k}$) from linearity of quantum evolutions that is in turn implied by unitarity of postulate (ii).

Everett settled (or at least bypassed) the ``collapse'' part of the problem with (iv) -- observer perceives the state of the rest of the Universe relative to his / her records. This is the essence of the Relative State Interpretation. 

However, from the standpoint of our quest for classical reality perhaps the most significant
and disturbing implication of (iv) is that quantum states do not exist -- at least not in the objective sense
to which we are used to in the classical world. The outcome of the measurement is typically {\it not} the preexisting state of the system, but one of the eigenstates of the measured observable.

Thus, whatever quantum state is, ``objective existence'' independent of what is known about it is clearly not one of its attributes. This malleability of quantum states clashes with the classical idea of what the state should be. Some even go as far as to claim that quantum states are simply a description
of the information that an observer has, and have essentially nothing to do with ``existence". 

I believe this denial of existence under any circumstances is going
too far -- after all, there are situations when a state can be found out, and the repeatability postulated by (iii) recognizes that its existence can be confirmed. But, clearly, (iv) limits the ``quantum existence'' of states to situations that are ``under the jurisdiction'' of postulate (iii) (or slightly more general situations where the preexisting density matrix
of the system commutes with the measured observable).

Collapse postulate (iv) complicates interpreting quantum formalism, as has been appreciated since Bohr and von Neumann \cite{11,59}. Therefore, at least before Everett, it was often cited as an indication of the ultimate insolubility of the ``quantum measurement problem''.
Yet, (iv) is hard to argue with -- it captures what happens in the laboratory measurements. 

To resolve the clash between the mathematical structure of quantum theory and our perception of what happens in the laboratory, in the real world measurements, one can accept 
-- with Bohr -- the primacy of our experience. The inconsistency of (iv) with the mathematical core of the quantum formalism -- superpositions of (i) and unitarity of (ii) -- can then be blamed
on the nature of the apparatus. According to the Copenhagen Interpretation the apparatus is classical, and,
therefore, not a subject to the quantum principle of superposition (which
follows from (i)). 
Measurements straddle the quantum - classical border, so they need not abide by the unitarity of (ii). Therefore, collapse can happen on the ``lawless'' quantum-classical border.

This quantum-classical duality posited by Bohr challenges the unification instinct of physicists.
One way of viewing decoherence is to regard einselection as a mechanism that accounts for effective classicality by suspending the validity of the quantum  
principle of superposition in a subsystem while upholding it for the composite system that includes the environment \cite{71,75}.

Everett's alternative to Bohr's approach was to abandon the literal
collapse and recognize that, once the observer is included
in the wavefunction, one can consistently interpret the consequences of such correlations. 
The right hand side of Eq. (1) contains all the possible outcomes, so the observer who records 
outcome \#17 perceives the branch of Universe that is consistent with that event reflected 
in his records. This view of the collapse is also consistent with repeatability of postulate (iii);  
re-measurement by the same observer using the same (non-demolition) device yields the same outcome.

Nevertheless, this relative state view of the quantum Universe suffers from a basic problem: the principle
of superposition (the consequence of axiom (i)) implies that the state of the system or of the apparatus
after the measurement can be written in infinitely many unitarily equivalent basis sets in the Hilbert
spaces of the apparatus (or of the observer's memory);
$$
\sum_k a_k \ket {s_k} \ket {A_k}=\sum_k a'_k \ket {s'_k} \ket {A'_k} = \sum_k a''_k \ket {s''_k} \ket {A''_k} =... \eqno(2)
$$
This is the {\it basis ambiguity} \cite{69}. It appears as soon as -- with Everett -- one eliminates
axiom (iv). The bases employed above are typically non-orthogonal, but in the Everettian relative state setting there
is nothing that would preclude them, or that would favor, e.g., the Schmidt basis of $\cS$ and $\cA$
(the orthonormal basis that is unique, provided that the absolute values of the {\it Schmidt coefficients} in such a {\it Schmidt decomposition} of an entangled bipartite state differ).

In our everyday reality we do not seem to be plagued by such basis ambiguity problems. So, in our
Universe there is something that (in spite of (i) and the egalitarian superposition principle it implies)
picks out preferred states, and makes them effectively classical. Axiom (iv) anticipates this.

Consequently, before there is an (apparent) collapse in the sense of Everett, a set of preferred states
-- one of which is selected by (or at the very least, consistent with) observer's records -- must be
chosen. There is nothing in the writings of Everett that would even hint that he was aware
of basis ambiguity and questions it leads to.

The next question concerns probabilities: how likely is it that, after I measure, my state
will be, say, $\ket {{\cal I}_{17}}$? Everett was keenly aware of this issue, and even believed that he
solved it by deriving Born's rule. In retrospect, it is clear that the argument he proposed -- as well as the arguments
proposed by his followers, including DeWitt \cite{21,22,20}, Graham \cite{22}, and Geroch \cite{31} who noted failure of Everett's original approach, and attempted to fix the problem --
did not accomplish as much as was hoped for, and did not amount to a derivation of Born's rule
(see \cite{48,49,37} for influential critical assessments).

In textbook versions of the quantum postulates probabilities are assigned
by another (Born's rule) axiom:

(v) {\it The probability $p_k$ of an outcome $\ket {s_k}$ in a measurement of a quantum system that was
previously prepared in the state $\ket \psi$ is given by $|\bk {s_k} \psi |^2$}.

Born's rule fits very well with Bohr's approach to quantum - classical transition (e.g.
with postulate (iv)). However, Born's rule is at odds with the spirit of the relative state
approach, or any approach that attempts (as we do) to deduce perception of the classical everyday reality starting from the quantum laws that govern our Universe. This does not mean that there is a mathematical inconsistency here: one can certainly
use Born's rule (as the formula $p_k= |\bk {s_k} \psi |^2$ is known) along with the relative state
approach in averaging to get expectation values and the reduced density matrix. 

Indeed, until the derivation of Born's rule in a framework of decoherence was proposed, 
decoherence practice relied on probabilities given by $p_k= |\bk {s_k} \psi |^2$. They enter whenever one assigns physical interpretation to reduced density matrices, a key tool of the decoherence theory. Everett's point was not that Born's rule is wrong, but,
rather, that it should be {\it derived} from the other quantum postulates, and we shall show how to do that. 

\section{Quantum Origin of Quantum Jumps}

To restate briefly the three problems identified above, we
need to derive the essence of the collapse postulates (iv)
and Born's rule (v) from our {\it credo} -- the core quantum postulates (o) - (iii).  Moreover, even when we accept relative
state origin of ``single outcomes'' and ``collapse'', we still need to justify emergence of the preferred basis that is
the essence of (iv). 

This issue (which in our summary of textbook axiomatics of quantum theory is a part of the collapse postulate) is so important that it is often captured by a separate postulate which declares that "Observables are Hermitian". This, in effect, means that the outcomes of measurements should correspond to orthogonal states in the Hilbert space. Furthermore, we should do it without appealing to Born's rule
-- without decoherence, or at least without its usual tools such as reduced density matrices that rely on Born's rule. Once we have preferred states, we will also have a set
of candidate {\it events}.  Once we have events we shall be able to pose questions about their probabilities.

The {\it preferred basis problem} was settled by the environment-induced superselection ({\it einselection}), 
usually regarded as a principal consequence of decoherence. This is discussed elsewhere \cite{69,70}.
Preferred pointer states and einselection are usually justified by appealing to decoherence. Therefore,
they come at a price that would have been unacceptable to Everett: decoherence and einselection
employ reduced density matrices and trace, and so their predictions are based on averaging, and thus, on probabilities -- on Born's rule.

Here we present an alternative strategy for arriving at preferred states that -- while not at odds with decoherence -- does not rely on the Born's rule-dependent tools of decoherence. Our overview of the origin of quantum jumps is brief. However, we direct the reader to references where different steps of that strategy are discussed in more detail. In short, we describe how one should go about doing the necessary physics, but we only sketch what needs to be done, and we do not do not explain all the details -- the requisite steps are carried out in the references we provide: Our discussion is meant as a guide to the literature, and not a substitute.

Decoherence done ``in the usual way'' (which, by the way, is a step in the right direction, in
understanding the practical and even many of the fundamental aspects of the quantum-classical
transition!) is not a good starting point in addressing the more fundamental aspects of the origins
of the classical. 

In particular, decoherence is not a good starting point for the derivation of Born's rule.
We have already noted the problem with this strategy: it courts circularity.
It employs Born's rule to arrive at the pointer states by using reduced density matrix which is obtained through trace -- i.e., averaging, which is where Born's rule is implicitly invoked (see e.g.~\cite{NC}). So, using decoherence to derive Born's rule is at best a consistency check. 

While I am most familiar with my own transgressions in this matter \cite{74}, this circularity also afflicts other approaches, including the proposal based on decision theory \cite{19,61,50}, as noted also by Refs. \cite{29,Synthese} among others. Therefore, one has to start the task from a different end. 

To get anywhere -- e.g., to define ``events'' essential in the introduction of
probabilities -- we need to show how the mathematical structure of quantum theory (postulates (o), (i)
and (ii) -- Hilbert space and unitarity) supplemented by the uncontroversial postulate (iii) (immediate repeatability, hence predictability) leads to preferred sets of states.

\subsection{Quantum Jumps from Quantum Core Postulates}

Surprisingly enough, deducing preferred states from ``quantum credo'' turns out to be simple.  
The possibility of repeated confirmation of an outcome is all that is needed to establish an effectively classical domain within the quantum Universe,
and to define events such as measurement outcomes.

One can accomplish this with minimal assumptions (``quantum core'' postulates (o) - (iii) on the above list) as described in
Ref. \cite{79,Z2013}. Here we review the basic steps. We assume that $\ket v$ and
$\ket w$ are among the possible repeatably accessible outcome states of ${\cal S}$.
$$ \ket v \ket {A_0} \Longrightarrow \ket v \ket {A_v} \  , \eqno(3a)$$
$$ \ket w \ket {A_0} \Longrightarrow \ket w \ket {A_w} \  . \eqno(3b)$$
So far, we have employed postulates (i) and (iii). The measurement, when repeated, would yield the same outcome, as the pre-measurement states have not changed. Thus, postulate (iii) is indeed satisfied. 

We now assume the process described by
Eq. (3) is fully quantum, so postulate (ii) -- unitarity of evolutions -- must also apply.
Unitarity implies that the overlap of the states before and after must be the same. Hence:
$$ \bk v w (1- \bk {A_v} {A_w})=0 \ . \eqno(4)$$
Our conclusions follow from this simple equation. There are two possibilities that depend on the overlap $\bk v w$. 

Suppose first that $\bk v w \neq 0$.
One is then forced to conclude that the measurement was unsuccessful since the state of ${\cal A}$ was unaffected by the process above. That is, the transfer
of information from ${\cal S}$ to ${\cal A}$ must have failed completely as in this case $\bk {A_v} {A_w}=1$ must hold. In particular, the apparatus can bear no imprint that distinguishes between states $\ket v$ and $\ket w$ that aren't orthogonal.

The other possibility, $\bk v w = 0$, allows for an arbitrary $\bk {A_v} {A_w}$, including a perfect
record, $\bk {A_v} {A_w}=0$. Thus, outcome states must be orthogonal if -- in accord with postulate (iii) -- they are to survive intact a
successful information transfer in general or a quantum measurement in particular, so that immediate re-measurement can yield the same result.

The same derivation can be carried out for ${\cal S}$ with a Hilbert space of
dimension ${\cal N}$ starting with a system state vector $\ket {\psi_{\cal S}}=\sum_{k=1}^{\cal N}\alpha_k \ket {s_k}$,
where (as before) -- {\it a priori} $\{\ket {s_k} \}$ need to be only linearly independent.

The simple reasoning above leads to a surprisingly decisive conclusion: orthogonality of the outcome states of the system is absolutely
essential for them to imprint even a minute difference on the state of any other system while retaining their
identity. The overlap $\bk v  w$ must be 0 {\it exactly} for $\bk {A_v} {A_w}$ to differ from unity. 

Imperfect or accidental information transfers (e.g., to the environment in course of decoherence) can also define
preferred sets of states providing that the crucial nondemolition demand of postulate (iii) is imposed on the unitary evolution responsible for the information flow.

A straightforward extension of the above derivation to where it can be applied not just to measured quantum systems (where nondemlition is a tall order), but to the measuring devices (where repeatability is essential) is possible~\cite{79,Z2013}. It is somewhat more demanding technically, as one needs to allow for mixed states and for decoherence in a model of a presumably macroscopic apparatus, but the conclusion is the same: Records maintained by the apparatus or repeatably accessible states of macroscopic but ultimately quantum systems must correspond to orthogonal subspaces of their Hilbert space.

It is important to emphasize that we are not asking for clearly distinguishable records (i.e., we are not demanding orthogonality of the states of the apparatus, $\bk {A_v} {A_w}=0$). 
Indeed, in the macroscopic case \cite{Z2013} one does not even ask for the state of the system to remain unchanged, but only for the outcomes of the consecutive measurements to be identical (i.e., the evidence of repeatability is in the outcomes). Still, even under these rather weak assumptions one is forced to conclude that {\it quantum states can exert distinguishable influences and remain unperturbed only when they are orthogonal}. To arrive at this conclusion we only used postulate (i) --
the fact that when two vectors in the Hilbert space are identical then physical states they correspond
to must also be identical.

{\begin{figure}[tb]
\begin{tabular}{l}
\vspace{-0.15in} 
\includegraphics[width=3.5in]{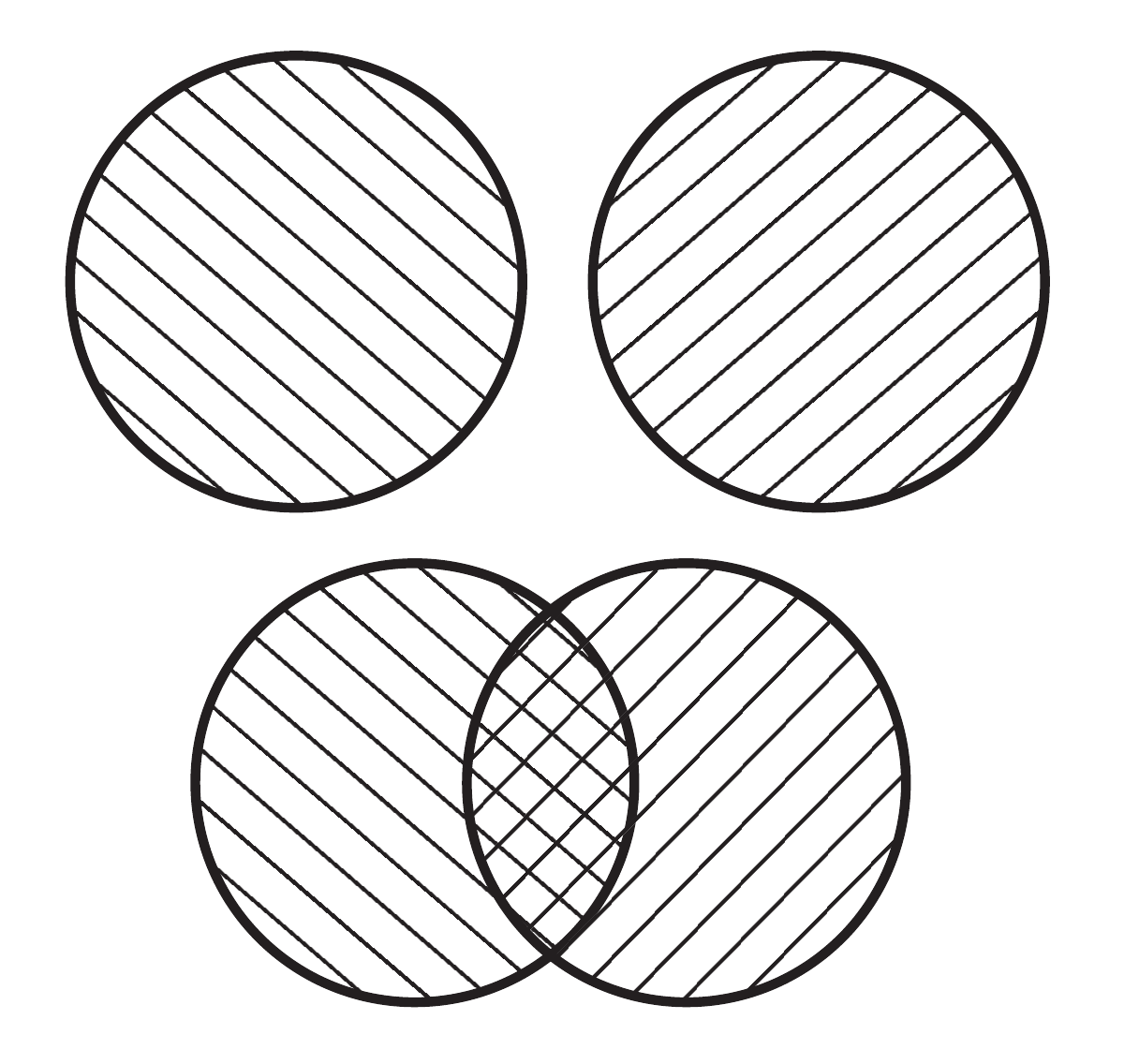}\\
\end{tabular}
\caption{The fundamental (pre-quantum) connection between distinguishability and repeatability of measurements. The two circles represent two states of the measured system. They correspond to two outcomes -- e.g., two properties of the underlying states (represented by two cross-hatchings). A measurement that can result in either outcome -- that can produce a record correlated with these two properties -- can be repeatable only when the two corresponding states (the two circles) do not overlap (case illustrated at the top). Repeatability is impossible without distinguishability: When two states overlap (case illustrated in the bottom), repetition of the measurement can always result in a system switching the state (and, thus, defying repeatability).
In the quantum setting this pre-quantum connection between repeatability and distinguishability leads to the derivation of orthogonality of repeatable measurement outcomes (and the two cross-hatchings can be thought of as two linear polarizations of a photon -- orthogonal on the top, but not below), but the basic intuition demanding distinguishability as a prerequisite for repeatability does not rely on quantum formalism.}
\label{Distinguishability}
\end{figure}

\subsection{Discussion}

Emergence of orthogonal outcome states is established above on the
foundation of very basic (and very quantum) assumptions. It leads one to conclude that observables are indeed associated with Hermitan operators. 

Hermitian observables are usually introduced in a very different manner -- they are the (postulated!) quantum versions of the familiar classical quantities. This emphasizes physical significance of their spectra (especially when they correspond to conserved quantities). Their orthogonal eigenstates emerge from the mathematics, once their Hermitian nature is assumed. Here we have deduced their Hermiticity by proving orthogonality of their eigenstates -- possible outcomes -- from the quantum core postulates by focusing on the effect of information transfer on the measured system.

The restriction to an orthogonal set of outcomes yields {\bf preferred basis}:
the essence of the collapse axiom (iv) need not be postulated! It follows from the
uncontroversial quantum core postulates (o)-(iii). 

We note that the preferred basis arrived at in this manner essentially coincides with
the basis obtained long time ago via einselection, \cite{69,70}. It is just that here we have arrived at this
familiar result without implicit appeal to Born's rule, which is essential if we want to take the next step,
and derive postulate (v).

We have relied on unitarity, so we did not derive the actual {\it collapse} of the wavepacket to a single outcome -- single event. Collapse 
is nonunitary, so one cannot deduce it starting from the quantum core that includes postulate (ii). However, we have accounted for one of the key collapse attributes:
the necessity of a symmetry breaking -- of the choice of a single {\it orthonormal} set of states from amongst various possible basis sets each of which can equally
well span the Hilbert space of the system -- follows from the core quantum postulates. This sets he stage for collapse -- for quantum jumps.

As we have already briefly noted, this reasoning can be extended \cite{Z2013} to when repeatably copied states belong to a macroscopic, decohering system (e.g., an apparatus pointer). In that case microstate {\it can} be perturbed by copying (or by the environment). What matters then is not the ``nondemolition'' of the microstate of the pointer, but persistence of the record its macrostate (corresponding to a whole collection of microstates) represents. To formulate this demand precisely one can rely on repeatability of copies: For instance, even though microstates of the pointer change upon readout due to the interaction with the environment, its macrostate should still represent the same measurement outcome -- it should still contain the same ``actionable information'' \cite{Z2013}. This more general discussion addresses also other issues (e.g., connection between repeatability, distinguishability, and POVM's, raised in {\bf FAQ \#4}) that arise in realistic settings (see Fig. 1 for the illustration of the key idea).

\section{Probabilities from Entanglement}

Derivation of events allows and even forces one to enquire about their probabilities or -- more specifically -- about the relation 
between probabilities of measurement outcomes and the initial pre-measurement state. 
As noted earlier, several past attempts at the derivation of Born's rule turned out
to be circular. Here we present key ideas behind a circularity-free approach.

We emphasize that our derivation of events does not rely on Born's rule. In particular, we have not attached any physical interpretation to the values of scalar products, and the key to our conclusions rested on whether the scalar product is (or is not) 0 or 1, or neither.

We now briefly review envariant derivation of Born's rule based on the symmetry of
entangled quantum states -- on {\bf entanglement - assisted invariance} or {\bf envariance}.
The study of envariance as a physical basis of Born's rule started with \cite{76,78,75}, and is now the focus of
several other papers (see e.g. Refs. \cite{53,5,H}). The key idea is illustrated in Fig. 2.

As we shall see, the eventual loss of coherence between pointer states
can be also regarded as a consequence of quantum symmetries of the states of systems entangled
with their environment. Thus, the essence of decoherence arises from symmetries of entangled states.
Indeed, some of the consequences of einselection  (including emergence of preferred states, as we have seen it in the previous section) can be studied without
employing the usual tools of decoherence theory (reduced density matrices and trace) that, for their physical significance, rely on Born's rule.

Decoherence that follows from envariance also allows one to justify additivity of probabilities, while the derivation of Born's rule 
by Gleason \cite{Gleason} assumed it (along with the other Kolmogorov's axioms of the measure-theoretic formulation of the foundations of probability theory, and with the Copenhagen-like setting). 
Appeal to symmetries leads to additivity also in the classical setting (as was noted already by Laplace: see \cite{40,32}. Moreover,
Gleason's theorem (with its rather complicated proof based on ``frame functions'' introduced especially for this purpose) provides no motivation why the measure he obtains should have any
physical significance -- i.e., why should it be regarded as probability. As illustrated in Fig. 2 and discussed below,
envariant derivation of Born's rule has a transparent physical motivation.

Additivity of probabilities is a highly nontrivial point. In quantum theory the overarching
additivity principle is the quantum principle of superposition. Anyone familiar with the double slit
experiment knows that probabilities of quantum states (such as the states corresponding to passing
through one of the two slits) do {\it not} add, which in turn leads to interference patterns. 

The presence of entanglement eliminates
local phases (thus suppressing quantum superpositions, i.e. doing the job of decoherence). This leads to additivity of probabilities of  events associated with preferred pointer states. 

\subsection{Decoherence, Phases, and Entanglement}

Decoherence is the loss of phase coherence between preferred states. It occurs when $\cS$ starts in a superposition of pointer states singled out by the interaction (represented below by the Hamiltonian ${{\bf H}_{\cS\cE}}$). As in Eq. (3), states of the system leave imprints -- become `copied' -- but now $\cS$ is `measured' by $\cE$, its environment:
$$
(\alpha \ket \uparrow + \beta \ket \downarrow)\ket {\varepsilon_0} { \stackrel {{\bf H}_{\cS\cE}} \Longrightarrow } \alpha \ket \uparrow \ket {\varepsilon_\uparrow}  + \beta \ket \downarrow \ket {\varepsilon_\downarrow} = \ket {\psi_{\cS\cE}}. \eqno(5)
$$
Equation (4) implied that the untouched states are orthogonal, $\bk \uparrow  \downarrow = 0$. Their 
superposition,
$\alpha \ket \uparrow + \beta \ket  \downarrow$ 
turns into an entangled $\ket {\psi_{\cS\cE}}$. Thus, neither $\cS$ nor $\cE$ alone have a pure state. This loss of purity signifies decoherence. One can still assign a mixed state that represents surviving information about $\cS$ to the system. 

Phase changes can be detected.: In a spin $\frac 1 2$--like $\cS$ $\ket \rightarrow = \frac {\ket \uparrow + \ket \downarrow} {\sqrt 2}$ is orthogonal to $\ket \leftarrow =\frac { \ket \uparrow - \ket \downarrow} {\sqrt 2}$.  Phase shift operator ${\bf u}^{\varphi}_\cS=\kb \uparrow \uparrow +  e^{\i \varphi}  \kb \downarrow \downarrow$ alters phase that distinguishes them: for instane, when $\varphi=\pi$, it converts $\ket  \rightarrow $ to $\ket \leftarrow$. In experiments ${\bf u}_\cS^\varphi$ would shift the interference pattern.

\hocom{In pure states phases matter; $\ket \rightarrow = \ket \uparrow + \ket \downarrow$ is orthogonal to $\ket \leftarrow = \ket \uparrow - \ket \downarrow$. One can adjust phases by acting, on $\cS$, with ${\bf u}_\cS^\phi=\kb \uparrow \uparrow +  e^{\i \phi}  \kb \downarrow \downarrow$. This phase shift operator converts $\ket  \rightarrow $ to $\ket \leftarrow$ when $\phi=\pi$. }

We assume perfect decoherence, $\bk {\varepsilon_\uparrow}{\varepsilon_\downarrow} = 0$: $\cE$ has a perfect record of pointer states. 
What information survives decoherence, and what is lost?

Consider someone who knows the initial 
pre-decoherence state, $\alpha \ket \uparrow + \beta \ket \downarrow$, and would like to make predictions about the decohered $\cS$. 
We now show that 
when $\bk {\varepsilon_\uparrow}{\varepsilon_\downarrow} = 0$ 
phases of $\alpha$ and $\beta$ no longer matter for $\cS$ -- phase $\varphi$ has no effect on {\it local} state od $\cS$, so measurements on $\cS$ cannot detect phase shift, as there is no interference pattern to shift.

Phase shift 
${\bf u}_\cS^\varphi \otimes {\bf 1}_\cE$ (acting on an entangled $\ket {\psi_{\cS\cE}}$) 
cannot have any effect on its local state because it
can be undone by ${\bf u}_{\cE}^{-\varphi}=\kb {\varepsilon_\uparrow} {\varepsilon_\uparrow} + e^{-\i \varphi} \kb {\varepsilon_\downarrow}{\varepsilon_\downarrow}$, a `countershift' acting on a distant $\cE$ decoupled from the system: 
$${\bf u}_{\cE}^{-\varphi}({\bf u}_\cS^{\varphi} \ket {\psi_{\cS\cE}})={\bf u}_{\cE}^{-\varphi}(\alpha \ket \uparrow \ket {\varepsilon_\uparrow}  + e^{\i \varphi}\beta \ket \downarrow \ket {\varepsilon_\downarrow})=\ket {\psi_{\cS\cE}} . \eqno(6) $$ 
Phases in $\ket {\psi_{\cS\cE}}$ can be changed in a faraway $\cE$
decoupled from but entangled with $\cS$. Therefore, 
 they can no longer influence local state of $\cS$. (This follows from quantum theory alone, but is essential for causality -- if they could, measuring $\cS$ would reveal this, enabling superluminal communication!)


Decoherence is caused by the loss of phase coherence. Superpositions 
decohere as $\ket \uparrow, \ket \downarrow$ are recorded by $\cE$. 
This is not because phases become ``randomized'' by interactions with $\cE$, as is sometimes said \cite{23}. Rather, 
they become delocalized: they lose significance for $\cS$ alone. They are a global property of the composite state -- they
no longer belong to $\cS$, so measurements on $\cS$ cannot distinguish states that started as superpositions with different phases for $\alpha, \beta$. 
Consequently, information about $\cS$ is lost -- it is displaced into correlations between $\cS$ and $\cE$, and local phases of $\cS$ become a global property -- global phases of the composite entangled state of $\cS\cE$. 

We have considered this information loss here without reduced density matrices, the usual decoherence tool. Our view of decoherence appeals to symmetry, invariance of $\cS$ -- {\it en}tanglement-assisted in{\it variance} or {\it envariance} under phase shifts of pointer state coefficients, Eq. (6). As $\cS$ entangles with $\cE$, its local state becomes invariant under transformations that could have affected it before. 


Rigorous proof of 
coherence loss uses quantum core postulates (o)-(iii) and relies on quantum {\it facts 1 -- 3}:

{\it  1. Locality: A unitary must act on a system to change its state.} State of $\cS$ that is not acted upon doesn't change even as other systems evolve (so ${\mathbf 1}_\cS \otimes (\kb {\varepsilon_\uparrow} {\varepsilon_\uparrow} + e^{-\i \varphi} \kb {\varepsilon_\downarrow}{\varepsilon_\downarrow})$
does not affect $\cS$ even when ${\cS\cE}$ are entangled, in $\ket {\psi_{\cS\cE}}$);

{\it 2. State of a system is all there is to predict measurement outcomes}; 

{\it 3. A composite state determines states of subsystems} (so local state of $\cS$ is restored when the state of the whole $\cS\cE$ is restored). 

{\it Facts} help characterize local states of entangled systems without using reduced density matrices. They follow from quantum theory: Locality is a property of interactions. the other two facts define the role and the relation of the quantum states of individual and composite systems in a way that does not invoke density matrices (to which we are not entitled in absence of Born's rule).
Thus, phase shift ${\bf u}_\cS^\varphi \otimes {\mathbf 1}_\cE=(\kb \uparrow \uparrow +  e^{\i \varphi}  \kb \downarrow \downarrow)  \otimes {\mathbf 1}_\cE$ 
acting on pure pre-decoherence state matters: 
measurement can reveal $\varphi$. In accord with facts 1 and 2, 
${\bf u}_\cS^\varphi$ changes $\alpha \ket \uparrow+ \beta \ket \downarrow$ into $\alpha \ket \uparrow+e^{\i \varphi}  \beta \ket \downarrow$.
However, the same ${\bf u}_\cS^\varphi$ acting on $\cS$ in an entangled state $\ket {\psi_{\cS\cE}}$ does not matter for $\cS$ alone, as it can be undone by ${\mathbf 1}_\cS \otimes (\kb {\varepsilon_\uparrow} {\varepsilon_\uparrow} + e^{-\i \varphi} \kb {\varepsilon_\downarrow}{\varepsilon_\downarrow})$, a countershift 
acting on a faraway, decoupled $\cE$. As the global $\ket {\psi_{\cS\cE}}$ is restored, by fact 3 the local state of $\cS$ is also restored even if $\cS$ is not acted upon (so that, by fact 1, it remains unchanged). 
Hence, local state of decohered $\cS$ that obtains from $\ket {\psi_{\cS\cE}}$ 
could not have changed to begin with, and so it cannot depend on phases of $\alpha, \beta$.

The only pure states invariant under such phase shifts (unaffected by decoherence) are pointer states. Resilience we saw, Eqs. (1)-(3), 
lets them preserve correlations.
For instance, entangled state of the measured system $\cS$ and the apparatus, $\ket {\psi_{\cS\cA}}$, Eq. (4),  decoheres as $\cA$ interacts with $\cE$:
$$(\alpha \ket \uparrow \ket {A_\uparrow}  + \beta \ket \downarrow \ket {A_\downarrow} ) \ket {\varepsilon_0}
{ \stackrel {{\bf H}_{\cA\cE}} \Longrightarrow }
\alpha \ket \uparrow \ket {A_\uparrow} \ket {\varepsilon_\uparrow}  + \beta \ket \downarrow \ket {A_\downarrow} \ket {\varepsilon_\downarrow} = \ket {\Psi_{\cS\cA\cE}} \eqno(7)$$
Pointer states $\ket {A_\uparrow}, \ket {A_\downarrow}$ of $\cA$ survive decoherence by $\cE$. They retain perfect correlation with $\cS$ (or an observer, or other systems) in spite of $\cE$, independently of the value of $\bk {\varepsilon_\uparrow} {\varepsilon_\downarrow}$. Stability under decoherence is -- in our quantum Universe -- a prerequisite for effective classicality: Familiar states of macroscopic objects also have to survive monitoring by $\cE$ and, hence, retain correlations.

\begin{figure*}[tb]
\begin{tabular}{l}
\vspace{-0.25in}
\includegraphics[width=3.6in]{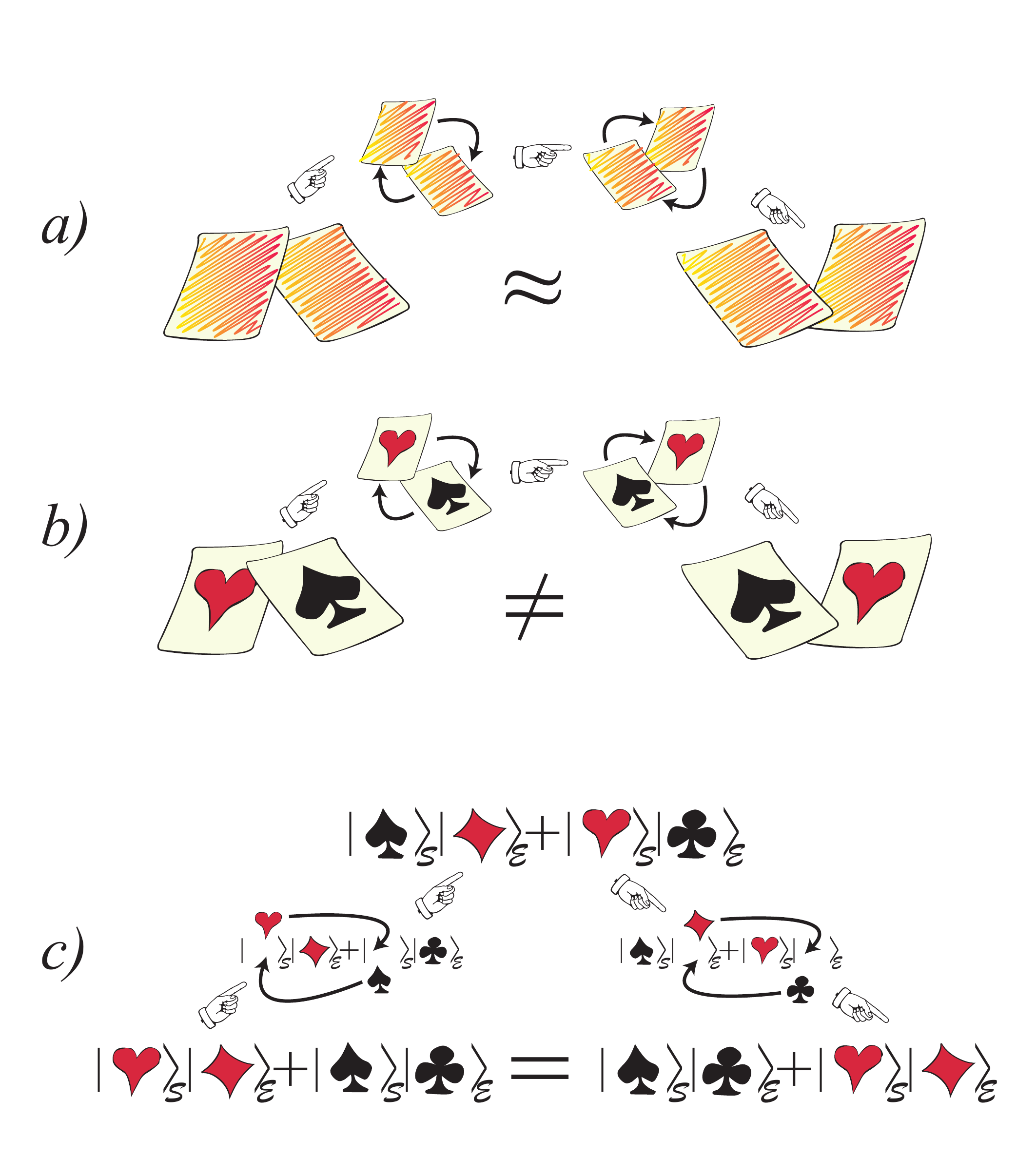}\\
\end{tabular}
\caption{Envariance -- {\it en}tanglement assisted in{\it variance} -- is a symmetry of entangled states. Envariance allows one to demonstrate Born's rule \cite{76,78,75} using a combination of an old intuition of Laplace \cite{40}
about invariance and the origins of probability and quantum symmetries of entanglement.
{\bf (a)} Laplace's {\it principle of indifference} (illustrated with playing cards) aims to establish
 symmetry using invariance under swaps.
A player who doesn't know face values of cards is  indifferent -- does not care -- if they are swapped before
he gets the one on the left. For Laplace, this indifference was the evidence of a (subjective) symmetry: 
It implied {\it equal likelihood} -- equal probabilities of the invariantly swappable
alternatives. For the two cards above, subjective probability $p_\spadesuit ={ \frac 1 2}$ would be inferred by someone who doesn't know their face value, but knows that one of them is a spade. 
When probabilities of a set of elementary events are provably equal, one
can compute probabilities of composite events and thus develop a theory of probability. Even the
additivity of probabilities can be {\it established} (see, e.g., Gnedenko, \cite{32}).
This is in contrast to Kolmogorov's measure-theoretic
axioms (which {\it include} additivity of probabilities). Above all, Kolmogorov's theory does not assign probabilities to elementary events (physical or otherwise), while envariant approach yields probabilities when symmetries of elementary events under swaps are known. 
{\bf (b)} The problem with Laplace's principle of indifference is its subjectivity.
The actual physical state of the system (the two cards) is altered by
the swap.  A related problem is that the assessment of indifference is based on ignorance: It
as was argued, e.g., by supporters of the relative frequency approach (regarded by many as more ``objective'' foundation of probability) that it is impossible to deduce anything 
(including probabilities) from ignorance. This is (along with subjectivity) was the reason why equal likelihood approach
is regarded with suspicion as a basis of probability in physics.
{\bf (c)} In quantum physics symmetries of entanglement can be used to deduce objective
probabilities starting with a known state. Envariance is the relevant symmetry.
When a pure entangled state of a system $\cS$ and another system we call ``an environment $\cE$'' (anticipating connections with decoherence) $
|\psi_{\cal SE}\rangle = \sum_{k=1}^N a_k |s_k\rangle |\varepsilon_k\rangle $ can be transformed
by $U_{\cal S}=u_{\cal S} \otimes {\bf 1}_{\cal E}$ acting solely on ${\cal S}$, but the effect of $U_{\cal S}$ can be undone by acting solely on ${\cal E}$ with an appropriately chosen $U_{\cal E}=
{\bf 1}_{\cal S} \otimes u_{\cal E}$, $U_{\cal E}|\eta_{\cal SE}\rangle  = ({\bf 1}_{\cal S} \otimes
u_{\cal E}) |\eta_{\cal SE}\rangle = |\psi_{\cal SE}\rangle $, it is envariant under $u_{\cal S}$. For such
composite states one can rigorously establish that the local state of ${\cal S}$ remains unaffected by
$u_{\cal S}$. Thus, for example, the phases of the coefficients in the Schmidt expansion
$|\psi_{\cal SE}\rangle = \sum_{k=1}^N a_k |s_k\rangle |\varepsilon_k\rangle $ are envariant, as the effect of $u_{\cal S}=\sum_{k=1}^N \exp(i \phi_k)|s_k\rangle \langle s_k| $ can be undone by a
{\it countertransformation} $u_{\cal E}=\sum_{k=1}^N \exp(-i \phi_k)|\varepsilon_k\rangle
\langle \varepsilon_k| $ acting solely
on the environment. This envariance of phases implies their
irrelevance for the local states -- in effect, it implies decoherence. Moreover, when the absolute values of the
Schmidt coefficients are equal 
a swap
$\ket \spadesuit \bra \heartsuit + \ket \heartsuit \bra \spadesuit$ in $\cS$ can be undone by a
`counterswap' $\ket \clubsuit \bra \diamondsuit + \ket \diamondsuit \bra \clubsuit$ in $\cE$.
So, as can be established more carefully \cite{78},  $p_\spadesuit = p_\heartsuit=\frac 1 2$
follows from the objective symmetry of such an entangled state. This proof of equal probabilities is based not on ignorance (as in Laplace's subjective `indifference') but on knowledge of the ``wrong property'' -- of the global observable that rules out (via quantum indeterminacy) any information about complementary local observables. When supplemented by simple counting, envariance leads to Born's rule also for unequal Schmidt coefficients \cite{76,78,75}.
}
\label{cards}
\end{figure*}

Decohered $\cS\cA$ is described by a {\it reduced density matrix}, 
$$ \rho_{\cS\cA} = \Tr_\cE \kb {\Psi_{\cS\cA\cE}}{\Psi_{\cS\cA\cE}} \ . \eqno(8a) $$ 
When $\bk {\varepsilon_\uparrow} {\varepsilon_\downarrow} = 0$, pointer states of $\cA$
retain correlations with the outcomes:
$$ \rho_{\cS\cA} 
= |\alpha|^2 \kb \uparrow \uparrow \kb {A_\uparrow} {A_\uparrow} + |\beta|^2 \kb \downarrow \downarrow \kb {A_\downarrow}{A_\downarrow} \eqno(8b) $$
Both $\uparrow$ and $\downarrow$ are present: There is no `literal collapse'.
We will use $\rho_{\cS\cA}$ to examine information flows. Thus, we will need probabilities of the outcomes.

Trace is a mathematical operation. However, regarding the reduced density matrix $\rho_{\cS\cA}$ as statistical mixture of its eigenstates -- states $\uparrow$ and $\downarrow$ and $A_\uparrow, A_\downarrow$ (pointer state) records -- relies on Born's rule, that allows one to view tracing as averaging. 
We didn't use it till Eq. (8) to avoid circularity. Now we derive $p_k=|\psi_k|^2$, Born's rule 
as we shall need it: We need to prove that the probabilities are indeed given by the eigenvalues $|\alpha|^2,  |\beta|^2$ of $\rho_{\cS\cA}$.  
This is the postulate (v), obviously crucial for relating quantum formalism to experiments. We want to deduce Born's rule from the quantum core postulates (o)-(iii). 

We note that this brief and somewhat biased discussion of the origin of decoherence is not a substitute for more complete presentations that do not (as we did, for good reasons in the present context) from employing usual tools of decoherence theory, including in particular reduced density matrices \cite{36,71,75}.

\subsection{Probabilities from Symmetries of Entanglement}

\hocom{We start our derivation of the Born's rule with the case of equal probabilities: When the Schmidt
coefficients are equal, symmetries of entanglement force one to conclude that the probabilities
must be also equal. The crux of the proof is that, after a swap on the system, the probabilities of
the swapped states must be equal to the probabilities of their new partners in the
Schmidt decomposition (which did not yet get swapped). But -- when the coefficients are equal -- a swap on
the environment restores the original states. So the probabilities must be the same as if the swap
never happened. These two requirements (that a swap exchanges probabilities, and that it does not
change them) can be simultaneously satisfied only when probabilities are equal.}

In quantum physics one seeks probability of measurement outcome starting from a known state of $\cS$ and ready-to-measure state of the apparatus pointer $\cA$. Entangled state of the whole is pure, so (at least prior to the decoherence by the environment) there is no ignorance in the usual sense. 

However, 
{\it envariance} in a guise slightly different than before (when it accounted for decoherence) implies that mutually exclusive outcomes have certifiably equal probabilities:
Suppose $\cS$ starts as $\ket \rightarrow = { \ket \uparrow + \ket \downarrow} $, so interaction with $\cA$ yields $ { \ket \uparrow \ket {A_\uparrow}  + \ket \downarrow \ket {A_\downarrow}}$, an {\it even} (equal coefficient) state. 
(Here and below we skip normalization to save on notation).

Unitary {\it swap} $ \kb \uparrow \downarrow + \kb \downarrow \uparrow$ permutes states in $\cS$:

\vspace{-.22cm}

\[
\tikz[baseline]{
            \node[fill=gray!20,anchor=base] (t1)
            {$|\uparrow \rangle$};
        } 
| {A_\uparrow} \rangle 
+
\tikz[baseline]{
            \node[fill=gray!20,anchor=base] (t2)
            {$|\downarrow \rangle$};
        }
|{A_\downarrow} \rangle
\quad \longrightarrow \quad
| \downarrow\rangle |{A_\uparrow} \rangle + |\uparrow  \rangle | {A_\downarrow}\rangle . \ \eqnum ((9a)
\]
\begin{tikzpicture}[overlay]
        \path[->] (t1) edge [bend left=40] (t2);
        \path[->] (t2) edge [bend left=40] (t1);
\end{tikzpicture}

\vspace{.2cm} 

\noindent 
After the swap $ \ket \downarrow $ is as probable as $\ket {A_\uparrow}$ was (and still is), and $\ket \uparrow$ as $\ket {A_\downarrow}$.  Probabilities in $\cA$ are unchanged 
(as $\cA$ is untouched) so $p_\uparrow$ and $p_\downarrow$ must have been swapped. To prove equiprobability we now 
swap records in $\cA$:


%
\[
|\downarrow\rangle \tikz[baseline]{
            \node[fill=gray!20,anchor=base] (t3)
            {$| {A_\uparrow}\rangle$};
        } 
+
|\uparrow \rangle
\tikz[baseline]{
            \node[fill=gray!20,anchor=base] (t4)
            {$|{A_\downarrow} \rangle$};
        }
\quad \longrightarrow \quad
|\downarrow\rangle |{A_\downarrow} \rangle| + |\uparrow \rangle | {A_\uparrow}\rangle . \ \eqnum ((9b)
\]
\begin{tikzpicture}[overlay]
        \path[->] (t3) edge [bend left=40] (t4);
        \path[->] (t4) edge [bend left=40] (t3);
\end{tikzpicture}
%


\noindent  Swap in $\cA$ restores pre-swap $ \ket \uparrow \ket {A_\uparrow}  + \ket \downarrow \ket {A_\downarrow}$ without touching $\cS$, so (by fact 3) the local state of $\cS$  is also restored(even though, by fact 1, it could not have been affected by the swap of Eq. (9a)).
Hence (by fact 2), all predictions about $\cS$, {\it including probabilities}, must be the same!
Probability of $\ket \uparrow$ and $ \ket \downarrow$, (as well as of $\ket {A_\uparrow} $ and $\ket {A_\downarrow}$) are exchanged yet unchanged. Therefore, 
they must be equal. Thus, in our two state case 
$p_\uparrow=p_\downarrow= \frac 1 2$. For $N$ envariantly equivalent alternatives, $p_k= \frac 1 N\ \forall k$.

Getting rid of phases beforehand was crucial:
Swaps in an isolated pure states will, in general, change the phases, and, hence, change the state.  For instance, $\ket \spadesuit + i \ket \heartsuit$, after a swap $\ket \spadesuit \bra \heartsuit + \ket \heartsuit \bra \spadesuit$, becomes
$i \ket \spadesuit + \ket \heartsuit$, i.e., is orthogonal to the pre-swap state. 

The crux of the proof of
equal probabilities was that the swap does not change anything {\it locally}. This can be established
for entangled states with equal coefficients but -- as we have just seen -- is simply not true for a pure
unentangled state of just one system.

In the real world the environment will become entangled (in course of decoherence)
with the preferred states of the system of interest (or with the preferred states of the apparatus pointer).
We have already seen how postulates (i) - (iii) lead to preferred sets of states. We have also pointed
out that -- at least in idealized situations -- these states coincide with the familiar pointer states
that remain stable in spite of decoherence. So, in effect, we are using the familiar framework of
decoherence to derive Born's rule. Fortunately our conclusions about decoherence can be reached without
employing the usual (Born's rule - dependent) tools of decoherence (reduced density matrix and trace). 

So far we have only explained how one can establish equality of probabilities for the
outcomes that correspond to Schmidt states associated with coefficients that differ at most by a phase. This is not yet Born's rule. However, it turns out that this is the hard part of the proof: Once such equality is established, a simple counting argument (a version of that employed in  \cite{74,19,61,50}) leads to the relation between probabilities and unequal coefficients  \cite{76,78,75}.

Thus, for an uneven state $\ket {\phi_{\cS\cA}}=\alpha\ket \uparrow \ket {A_\uparrow}  + \beta \ket \downarrow \ket {A_\downarrow}$ swaps on $\cS$ and $\cA$ yield $\beta \ket \uparrow \ket {A_\uparrow}  + \alpha \ket \downarrow \ket {A_\downarrow}$, and not the pre-swap state, so $p_\uparrow$ and $p_\downarrow$ are not equal. However, uneven 
case reduces to equiprobability via {\it finegraining}, so envariance, Eq. (8), yields Born's rule, 
$p_{s|\psi}=|\bk s \psi|^2$, 
in general.

To see how, we take $\alpha \propto \sqrt \mu, ~ \beta \propto \sqrt \nu $, where $\mu, \nu$ are natural numbers (so the squares of $\alpha$ and $\beta$ are commensurate). To finegrain, we change the basis; $\ket {A_\uparrow}=\sum_{k=1}^\mu \ket {a_{ k}}/\sqrt \mu$, and $ \ket {A_\downarrow}=\sum_{k=\mu+1}^{\mu+\nu} \ket {a_{ k}}/\sqrt \nu$, in the Hilbert space of $\cA$:
\vspace{-0.1cm}
$$
 \ket {\phi_{\cS\cA} } \propto \sqrt \mu ~ \ket \uparrow \ket {A_\uparrow} + \sqrt \nu ~ \ket \downarrow \ket {A_\downarrow} = $$
\vspace{-0.5cm}
 $$
= \sqrt \mu ~ \ket \uparrow \sum_{k=1}^\mu \ket {a_{ k}}/\sqrt \mu + \sqrt \nu ~ \ket \downarrow\sum_{k=\mu+1}^{\mu+\nu} \ket {a_k}/\sqrt \nu \ . 
\eqno(10a) 
$$ 
We simplify, and imagine an environment decohering $\cA$ in a new orthonormal basis. That is, $\ket {a_k}$ correlate with $\ket {e_k}$ so that;
$$
\ket {\Phi_{\cS\cA\cE} } \propto\sum_{k=1}^\mu \ket {\uparrow {a_k}} \ket {e_k}+ \sum_{k=\mu+1}^{\mu+\nu}\ket {\downarrow {a_k} } \ket {e_k} 
\eqno(10b)
$$
as if $\ket {a_k}$ were the preferred pointer states decohered by the environment so that $\bk {{e_k}} {{e_l}} = \delta_{kl}$. 

Now swaps of $\ket {\uparrow {a_k}}$ with $\ket {\downarrow {a_k} } $ can be undone by counterswaps of the corresponding $\ket {e_k}$'s.
Counts of the finegrained equiprobable $(p_k=\frac 1 {\mu + \nu})$ alternatives labelled with $\uparrow$ or $\downarrow$ lead to Born's rule:
$$ p_\uparrow = \frac \mu {\mu + \nu} =|\alpha|^2, \ \ \ \ p_\downarrow = \frac  \nu {\mu + \nu} = |\beta|^2 . \eqno(11)$$
Amplitudes `got squared' as a result of Pythagoras' theorem (euclidean nature of Hilbert spaces). The case of incommensurate $|\alpha|^2$ and $ |\beta|^2$ can be settled by an appeal to continuity of probabilities as functions of state vectors. 

\subsection{Discussion}

In physics textbooks Born's rule is a postulate. Using entanglement we derived it here from the quantum core axioms. Our reasoning was purely quantum: Knowing a state of the composite classical system means knowing state of each part. There are no entangled classical states, and no objective symmetry to deduce classical equiprobability, the crux of our derivation. Entanglement -- made possible by the tensor structure of composite Hilbert spaces, introduced by the composition postulate (o) -- was key. Appeal to symmetry -- subjective and suspect in the classical case -- becomes rigorous thanks to objective envariance in the quantum case. 
Born's rule, introduced by textbooks as postulate (v), follows. We also note that envariance has been successfully tested in several recent experiments \cite{Laflamme,Ebrahim,Deffner,Ferrari}.

Relative frequency approach (found in many probability texts) starts with ``events''. It has not led to successful derivation of Born's rule.
We used entanglement symmetries to identify equiprobable alternatives. However, by employing envariance one can also deduce frequencies of events by considering $M$ repetitions (i.e., $(\alpha\ket \uparrow \ket {A_\uparrow}  + \beta \ket \downarrow \ket {A_\downarrow})^{\otimes M}$) of an experiment, and deduce departures that are also expected when $M$ is finite. Moreover, one can even show the inverse of Born's rule. That is, one can demonstrate that the amplitude should be proportional to the square root of frequency \cite{Zurek11}.

As the probabilities are now in place, one can think of quantum statistical physics. One could establish its foundations using probabilities we have just deduced. But there is an even simpler and more radical approach \cite{DefZur,Zurek18a} that arrives at the microcanonical state without the need to invoke ensembles and probabilities. Its detailed explanation is beyond the scope of this section, but the basic idea is to regard an even state of the system entangled with its environment as the microcanonical state. This is a major conceptual simplification of the foundations of statistical physics: One can get rid of the artifice of invoking infinite collections of similar systems to represent a state of a single system in a manner that allows one to deduce relevant thermodynamic properties. 
\section{Quantum Darwinism, Classical Reality, and Objective Existence}

\hocom{Monitoring of the system by the environment (the process responsible for
decoherence) will typically leave behind multiple copies of its pointer states in $\cE$. Pointer states are favored -- only states that can survive decoherence can produce information theoretic progeny in this manner \cite{42,43}. Therefore, only information about
pointer states can be recorded redundantly. States that can survive decoherence
can use the same interactions that are responsible for einselection to proliferate information about
themselves throughout the environment.}

{\begin{figure}[tb]
\begin{tabular}{l}
\vspace{-0.15in} 
\includegraphics[width=3.5in]{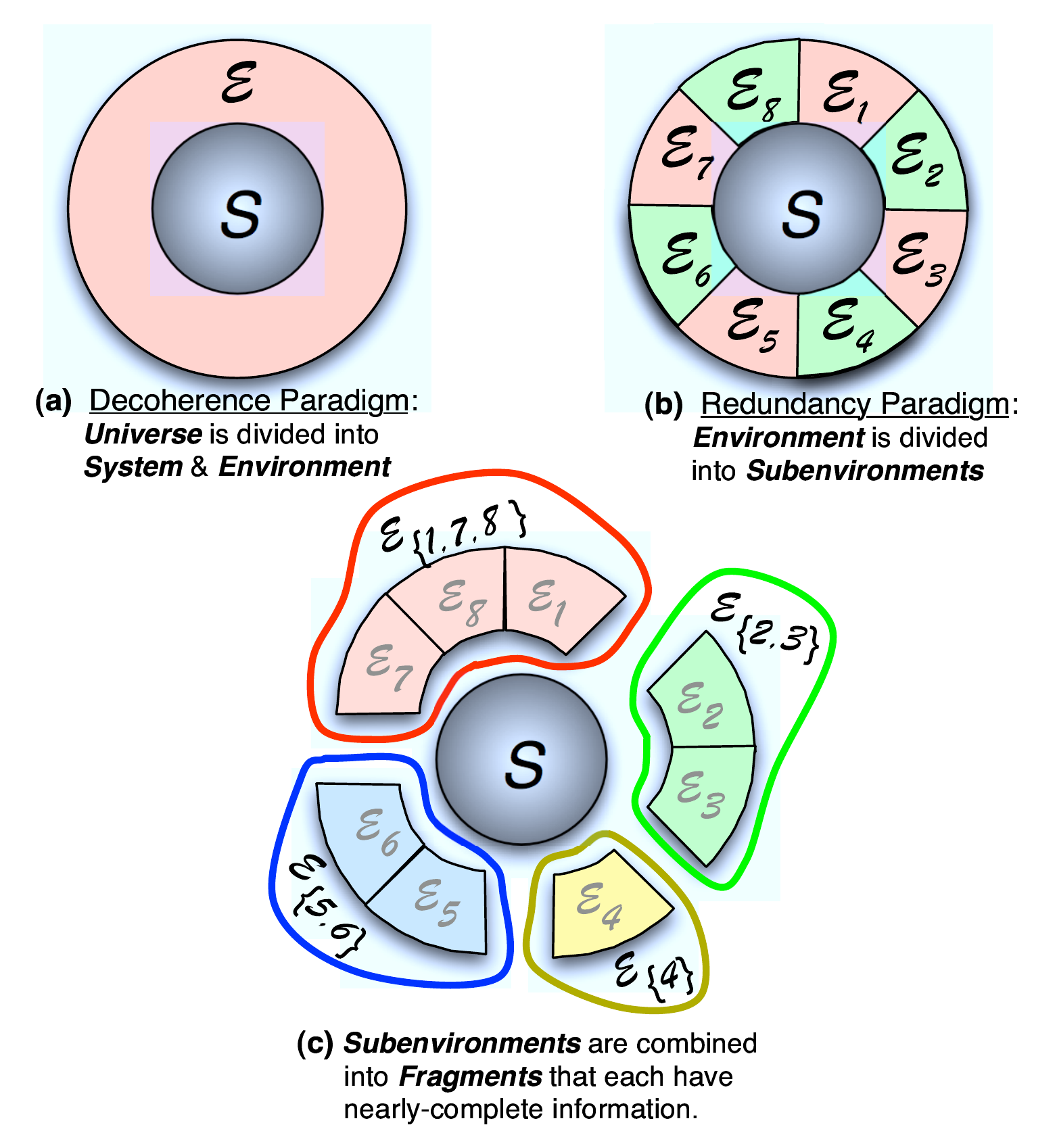}\\
\end{tabular}
\caption{Quantum Darwinism recognizes that environments consist of many subsystems, and that observers acquire information about system of interest $\cS$ by intercepting copies of its pointer states deposited in $\cE$ as a result of decoherence.
}
\label{EnvSubdivision}
\end{figure}

Quantum Darwinism \cite{75,Z00} recognizes that observers use the environment as
a communication channel to acquire information about pointer states indirectly, leaving the system of interest untouched
and its state unperturbed. Observers can
find out the state of the system without endangering its existence (which would be inevitable in direct measurements). Indeed, the reader of this text is -- at this very moment -- intercepting
a tiny fraction of the photon environment by his eyes to gather all of the information he needs. 

This is
how virtually all of our information is acquired. A direct measurement is not what we do. Rather, we
count on redundancy, and settle for information that exists in many copies. This is how objective existence -- cornerstone of classical reality -- arises in the quantum world. 


\hocom{ The existence of redundant copies of pointer states implies that observables which do not commute with
the pointer observable are inaccessible. The simplest model of quantum Darwinism that illustrates this
is a somewhat contrived arrangement of many ($N$) target qubits that constitute subsystems
of the environment interacting via a {\it controlled not} ({\tt c-not}) with a single control qubit $\cS$. As time goes on,
consecutive target qubits become imprinted with the state of the control $\cS$:
$$(a\ket 0+ b\ket 1)\otimes \ket {0_{\varepsilon_1}} \otimes \ket {0_{\varepsilon_2}}\dots \otimes \ket {0_{\varepsilon_N}} \Longrightarrow $$
$$
(a\ket 0\otimes \ket {0_{\varepsilon_1}} \otimes \ket {0_{\varepsilon_2}} + b\ket 1\otimes \ket {1_{\varepsilon_1}} \otimes \ket {1_{\varepsilon_2}})\dots \otimes \ket {0_{\varepsilon_N}} \Longrightarrow
$$
$$
a\ket 0\otimes \ket {0_{\varepsilon_1}} \otimes \dots \otimes \ket {0_{\varepsilon_N}} + b\ket 1\otimes \ket {1_{\varepsilon_1}} \dots \otimes \ket {1_{\varepsilon_N}} \ .  $$
This simple dynamics creates multiple records of the logical basis ``pointer'' states of the system in
the environment. The existence of the preferred pointer basis that is untouched by the interaction is
essential. As we have seen earlier, this is possible -- such quantum jumps emerge from the purely
quantum core postulates (o) - (iii).}

\subsection{Mutual Information in Quantum Correlations}

To develop theory of quantum Darwinism we need to quantify information between fragments of the environment and the system.
Mutual information is a convenient tool that we shall use for this purpose.

The mutual information between the system $\cS$ and a fragment $\cF$ (that will play the role opf the apparatus $\cA$ of Eq. (8) in the discussion above) can be computed using the density matrices of the systems of interest using their von Neumann entropies $H_X=-\Tr \rho_x \lg \rho_X$; 
$$ I(\cS : \cF)=H_\cS+ H_{\cF} - H_{\cS, \cF}=-(|\alpha|^2\lg|\alpha|^2+|\beta|^2\lg|\beta|^2) \eqno(12)$$
We have used the density matrices of the $\cS$ and $\cA$ (as a ``stand-in'' for $\cF$) from Eq. (8) to obtain the specific value of mutual information above.

We already noted the special role of the pointer observable. It is stable and, hence, it leaves behind
information-theoretic progeny -- multiple imprints, copies of the pointer states -- in the environment.
By contrast, complementary observables 
are destroyed by the interaction with a single subsystem of $\cE$. They can in principle still be accessed,
but only when {\it all} of the environment is measured. Indeed, because we are dealing with a quantum
system, things are much worse than that: The environment must be measured in
precisely the right (typically global) basis to allow for such a reconstruction. Otherwise, the accumulation of errors over multiple
measurements will lead to an incorrect conclusion and re-prepare the state and environment, so that it is
no longer a record of the state of ${\cal S}$, and phase information is irretrievably lost.

\hocom{As each environment qubit is a perfect copy of $\cS$, redundancy in this simple example is eventually
given by the number of fragments -- that is, in this case, by the number of the environment qubits -- that
have (more or less) complete information about $\cS$. In this simple case there is no reason to
define redundancy in a more sophisticated manner.  Such a need arises in more realistic
cases when the analogues of {\tt c-not}'s are imperfect.}

\subsection{Objective Reality form Redundant Information}

Quantum Darwinism was introduced relatively recently. Previous studies of the records ``kept'' by the environment were focused on its effect on the state of the system, and not on their utility. Decoherence is a case in point, as are some of the studies of the decoherent histories approach \cite{GMH,JJH}.  
The exploration of quantum Darwinism in specific models has started at he beginning of this millenium \cite{8,9,10,42,43}.
We do not intend to review all of the results obtained to date in detail.
The basic conclusion of these studies is, however, that the dynamics responsible for decoherence is
also capable of imprinting multiple copies of the pointer basis on the environment. Moreover, while
decoherence is always implied by quantum Darwinism, the reverse need not be true.  One can
easily imagine situations where the environment is completely mixed, and, thus, cannot be used as
a communication channel, but would still suppress quantum coherence in the system.

For many subsystems, $\cE=\bigotimes_k \cE^{(k)}$, the initial state $(\alpha \ket \uparrow 
+ \beta \ket \downarrow) 
\ket { {\varepsilon^{(1)}_0} {\varepsilon^{(2)}_0} {\varepsilon^{(3)}_0}...}$ evolves into a ``branching state'';
$$ \ket {\Upsilon_{\cS\cE}} = \alpha \ket \uparrow \ket { {\varepsilon^{(1)}_\uparrow} {\varepsilon^{(2)}_\uparrow}{\varepsilon^{(3)}_\uparrow}... } + \beta \ket \downarrow \ket { {\varepsilon^{(1)}_\downarrow} {\varepsilon^{(2)}_\downarrow}{\varepsilon^{(3)}_\downarrow}...} \eqno(13)$$
Linearity assures all branches persist: collapse to one outcome is not in the cards.
However, large $\cE$ can disseminate information about the system. The state $\ket {\Upsilon_{\cS\cE}}$ represents many records inscribed in its fragments, collections of subsystems of $\cE$ (Fig. 3). 
This means that the state of $\cS$ can be found out by many, independently, and indirectly---hence, without disturbing $\cS$. This is how symptoems of objective existence arises in our quantum world.

An environment fragment $\cF$ can act as apparatus with a (possibly incomplete) record of $\cS$. When $\cE \backslash \cF$ (`the rest of the $\cE$') is traced out, $\cS\cF$ decoheres, and the reduced density matrix describing joint state of $\cS$ and $\cF$ is:
$$ \rho_{\cS\cF} = \Tr_{\cE \backslash \cF} \kb {\Psi_{\cS\cE}}{\Psi_{\cS\cE}}= |\alpha|^2 \kb \uparrow \uparrow \kb {F_\uparrow} {F_\uparrow} + |\beta|^2 \kb \downarrow \downarrow \kb {F_\downarrow}{F_\downarrow} \eqno(14) $$
When $\bk {F_\uparrow} {F_\downarrow} =0$, $\cF$ contains perfect record of the preferred states of the system. 
In principle, each subsystem of $\cE$ may be enough to reveal its state, but this is unlikely. Typically, one must collect many subsystems of $\cE$ into $\cF$ to find out about $\cS$.

The redundancy of the data about pointer states in $\cE$ determines how many times the same information can be independently extracted---it is a measure of objectivity. 
The key question of quantum Darwinism is then: {\it How many subsystems of $\cE$---what fraction of $\cE$---does one need to find out about $\cS$?}. The answer is provided by the mutual information $I(\cS : \cF_f)=H_\cS + H_{\cF_f} - H_{\cS \cF_f}$, information about $\cS$ available from $\cF_f$, 
fraction $f= \frac { \sharp \cF } { \sharp \cE }$ of $\cE$ (where $\sharp \cF$ and $ \sharp \cE$ are the numbers of subsystems). 

In case of perfect correlation a single subsystem of $\cE$ would suffice, as $I(\cS : \cF_f)$ jumps to $H_\cS$ at $f=\frac 1 {\sharp \cE}$. The data in additional subsystems of $\cE$ are then redundant.
Usually, however, larger fragments of $\cE$ are needed to find out enough about $\cS$. Red plot in Fig. 4 illustrates this: $I(\cS : \cF_f)$ still approaches $H_\cS$, but only gradually. The length of this plateau can be measured in units of $f_\delta$, the initial rising portion of $I(\cS : \cF_f)$. It is defined with the help of the {\it information deficit} $\delta$ observers tolerate:
$$ I(\cS : \cF_{f_\delta}) \ge (1- \delta) H_\cS \eqno(15) $$
Redundancy is the number of such records of $\cS$ in $\cE$:
$$ {\cal R_\delta} = 1 / f_\delta \eqno(16)$$
$ {\cal R_\delta}$ sets the upper limit on how many observers can find out the state of $\cS$ from $\cE$ independently and indirectly. In models \cite{42,43,8,9,10,RiedelZ10,RiedelZ12,ZwolakRZ14} (especially photon scattering analyzed extending decoherence model of Joos and Zeh \cite{JZ}) $\cal R_\delta$ is huge \cite{RiedelZ10,RiedelZ12,ZwolakRZ14} and depends on $\delta$ only weakly (logarithmically). 


This is `quantum spam': $\cal R_\delta$ imprints of pointer states are broadcast through the environment. Many observers can access them independently and indirectly, assuring objectivity of pointer states of $\cS$. Repeatability is key: States must survive copying to produce many imprints. 

\subsection{Discussion}

Our discussion of quantum jumps shows when, in spite of the no-cloning theorem \cite{65,24}, repeatable copying is possible. Discrete preferred states set the stage for quantum jumps.
Copying yields branches of records inscribed in subsystems of $\cE$. Initial superposition yields superposition of branches, Eq. (13), so there is no literal collapse.
However, fragments of $\cE$ can reveal only one branch (and not their superposition). Such evidence will suggest `quantum jump' from superposition to a single outcome, in accord with (iv).

Not all environments are good in this role of a witness. Photons excel: They do not interact with the air or with each other, faithfully passing on information. Small fraction of photon environment usually reveals all we need to know. Scattering of sunlight quickly builds up redundancy: a 1$\mu$ dielectric sphere in a superposition of 1$\mu$ size increases ${\cal R}_{\delta=0.1}$ by $ \sim 10^8$ every microsecond \cite{RiedelZ10, RiedelZ12}. Mutual information plot illustrating this case is shown in Fig. 5.

{\begin{figure}[tb]
\begin{tabular}{l}
\vspace{-0.15in} 
\includegraphics[width=3.6in]{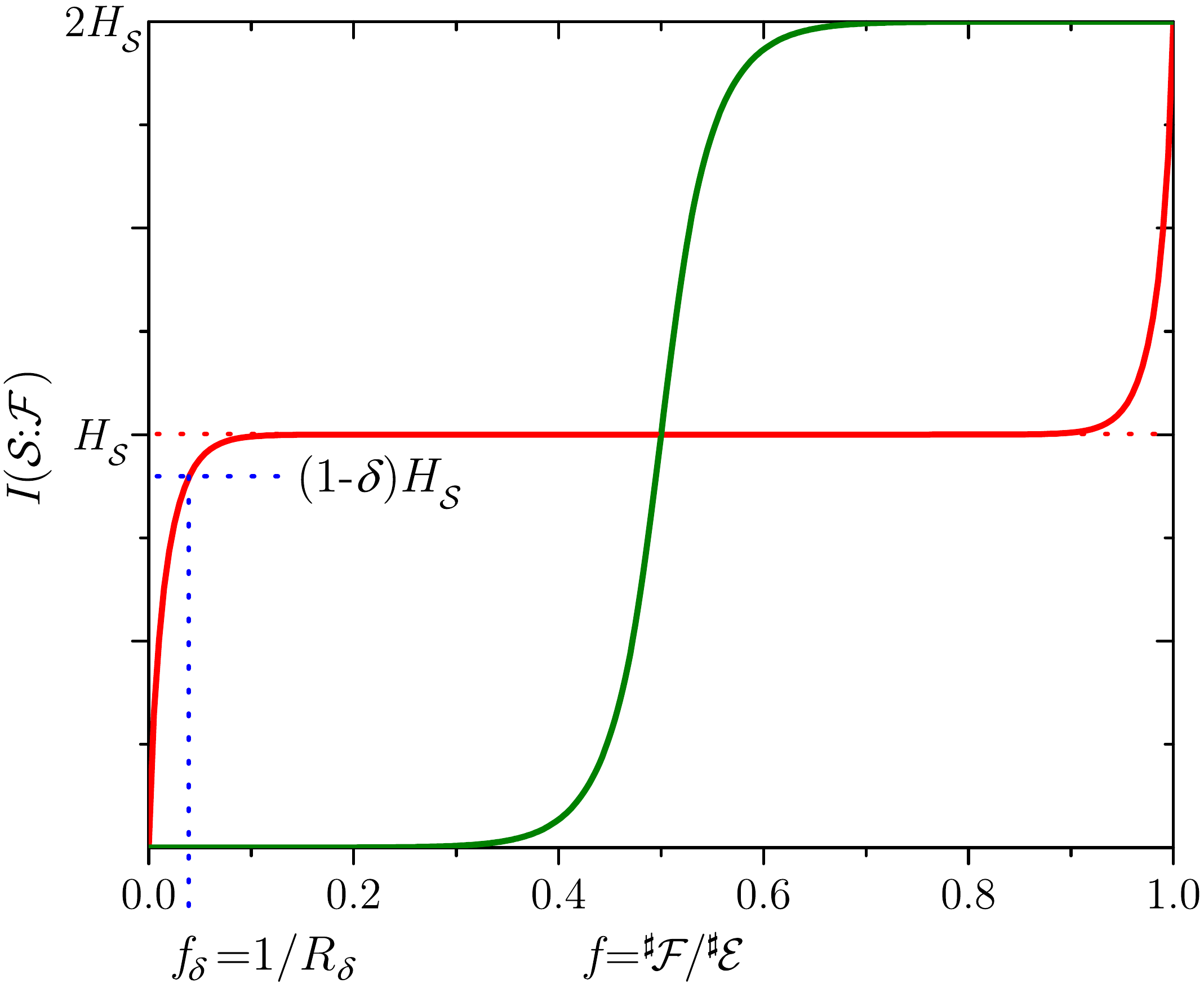}\\
\end{tabular}
\caption{Information about the system contained in a fraction $f$ of the environment. Red plot shows a typical $I(\cS : \cF_f)$ established by decoherence. Rapid rise means that nearly all classically accessible information is revealed by a small fraction of $\cE$. It is followed by a plateau: additional fragments only confirm what is already known. Redundancy ${\cal R_\delta} = 1 / f_\delta$ is the number of such independent fractions. Green plot shows $I(\cS : \cF_f)$ for a random state in the composite system ${\cS\cE}$.
}
\label{RedPIP}
\end{figure}

Air is also good in decohering, but its molecules interact, scrambling acquired data. Objects of interest scatter both air and photons, so both acquire information about position, and favor similar localized pointer states.

Quantum Darwinism shows why it is so hard to undo decoherence \cite{ZwolakZ}. Plots of mutual information $I(\cS : \cF_f)$ for initially pure $\cS$ and $\cE$ are antisymmetric (see Fig. 4) around $f= \frac 1 2$ and $H_\cS$ \cite{8}. Hence, a counterpoint of the initial quick rise at $f \le f_\delta$ is a quick rise at $f \ge 1 - f_\delta$, as last few subsystems of $\cE$ are included in the fragment $\cF$ that by now contains nearly all $\cE$. 
This is because an initially pure $\cS \cE$ remains pure under unitary evolution, so $H_{\cS \cE}=0$, and $I(\cS : \cF_f)|_{f=1}$ must reach $2 H_{\cS}$. Thus, a measurement on {\it all} of $\cS \cE$ could confirm its purity in spite of decoherence caused by $\cE \backslash \cF$ for all $f \le 1- f_\delta$. 
However, to verify this one has to intercept and measure all of $\cS\cE$ in a way that reveals pure state $\ket {\Upsilon_{\cS\cE}}$, Eq. (13). Other measurements destroy phase 
information. So, undoing decoherence is in principle possible, but the required resources and foresight preclude it.

In quantum Darwinism decohering environment acts as an amplifier,
inducing branch structure of $\ket {\Upsilon_{\cS\cE}}$ distinct from typical states in the Hilbert space of $\cS\cE$: $I(\cS : \cF_f)$ of a random state is given by the green plot in Fig. 4, with no plateau or redundancy. 
Antisymmetry means that $I(\cS : \cF_f)$ `jumps' at $f= \frac 1 2$ to $2 H_{\cS}$. 

Environments that decohere $\cS$, but scramble information because of interactions between  subsystems (e.g., air) eventually approach such random states. Quantum Darwinism is possible only when information about $\cS$ is preserved in fragments of $\cE$, so that it can be recovered by observers. There is no need for perfection: Partially mixed environments or imperfect measurements correspond to noisy communication channels: their capacity is depleted, but we can still get the message \cite{ZwolakQZ1, ZwolakQZ2}.

\begin{figure*}[tb]
\begin{tabular}{l}
\includegraphics[width=6.5in]{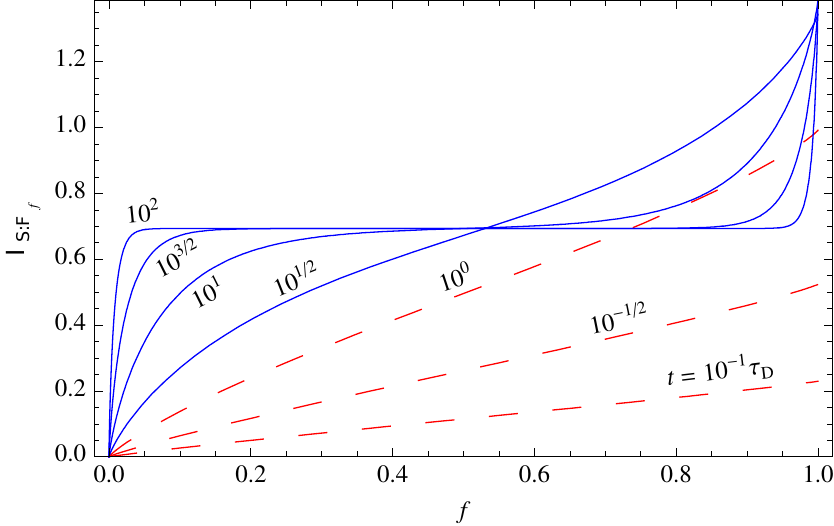}\\
\end{tabular}
\caption{The quantum mutual information $I(\cS : \cF_f )$ vs. fragment size $f$ at different elapsed times for an object illuminated by a point-source black-body radiation \cite{RiedelZ12}.
Individual curves are labeled by the time $t$ in units of the decoherence time $\tau_D$.
For $t \le\tau_D$ (red dashed lines), the information about the system available in the environment is low.  The linearity in $f$ means each piece of the environment contains new, independent information.  For $t>\tau_D$ (blue solid lines), the shape of the partial information plot indicates redundancy; the first few pieces of the environment increase the information, but additional pieces only confirm what is already known. 
\hocom{ The remaining information (i.e. above the plateau) is highly ``encrypted'' in the global state, in the sense that it can only read by capturing almost all of $\Env$ and measuring $\cS\cE$ in the right way. \\
(b) For isotropic illumination, the same time-slicing is used as in (a) but there is greatly decreased mutual information because the directional photon states are ``full'' and cannot store more information about the state of the object.}}
\label{illumination}
\end{figure*}

Quantum Darwinism settles the issue of the origin of classical reality by accounting for all of the operational symptoms of objective existence
in a quantum Universe: A single quantum state cannot be found out through a direct
measurement. However, pointer states usually leave multiple records in the environment. Observers can use these
records to find out the (pointer) state of the system of interest. Observers can afford to destroy photons while
reading the evidence -- the existence of multiple copies implies that other observers can access
the information about the system indirectly and independently, and that they will all agree about
the outcome.  This is how objective existence arises in our quantum world. 

There has been significant progress in the study of the acquisition and dissemination of the information by the environments \cite{QD}. More detailed discussion of the results obtained in these papers is, unfortunately, beyond the scope of our brief review.


\section{Discussion: Frequently Asked Questions}

The subject of this paper has a long history. As a result, there are different ways of talking, thinking, and writing about it. It is almost as if different points of view have developed different languages. As a result, one can find it difficult to understand the ideas, as one often has to learn ``the other language'' used to discuss the same problem. This is further complicated by the fact that all of these languages use essentially the same words, but charged with a very different meanings. Concepts like ``existence'', ``reality'', or ``state'' are good examples.

The aim of this section is to acknowledge this problem and to deal with it to the extent possible within the framework of a brief guide. We shall do that in a way inspired by modern approach to languages (and to travel guides): Rather than study vocabulary and grammar, we shall use ``conversations'' based on a few ``frequently asked questions''. 
The hope is that this exercise will provide the reader with some useful hints of what is meant by certain phrases. This is very much in the spirit of the ``travel guide'', where a collection of frequently used expressions is often included.

{\bf FAQ~\#1}: {\it What is the difference between ``decoherence'' and ``einselection''?}

Decoherence is the process of the {\it loss of phase coherence} caused by the interaction between the system and the environment. Einselection is an abbreviation of ``environment - induced superselection'', which designates {\it selection of preferred set of pointer states} that are immune to decoherence. Decoherence will often (but not always) result in einselection. For instance, interaction that commutes with a certain observable of a system will preserve eigenstates of that {\it pointer observable}, {\it pointer states} that are einselected, and do {\it not} decohere. By contrast, superpositions of such pointer states will decohere. This picture can be (and generally will be) complicated by the evolution induced by the Hamiltonian of the system, so that perfect pointer states will not exist, but approximate pointer states will be still favored -- will be much more stable then their superpositions. There are also cases when there is decoherence, but it treats all the states equally badly, so that there is no einselection, and there are no pointer state. Perfect depolarizing channel~\cite{NC} is an example of such decoherence that does not lead to einselection. Section III of this paper emphasizes the connection between predictability and einselection, and leads to a derivation of preferred states that does not rely on Born's rule.

{\bf FAQ~\#2}: {\it Why does axiom (iv) conflict with ``objective existence'' of quantum states?} 

The criterion for objective existence used here is pragmatic and operational: Finding out a state without prior knowledge is a necessary condition for a state to objectively exist~\cite{75,42,43,8,9,10}. Classical states are thought to exist in this sense. Quantum states do not: Quantum measurement yields an outcome -- but, according to axiom (iv), this is one of the eigenstates of the measured observable, and not a preexisting state of the system. Moreover, according to axiom (iii) (or collapse part of (iv)) measurement re-prepares the system in one of the eigenstates of the measured observable. A sufficient condition for objective existence is the ability of many observers to independently find out the state of the system without prior knowledge, and to agree about it. Quantum Darwinism makes this possible.

{\bf FAQ~\#3}: {\it What is the relation between the preferred states derived using their predictability (axiom (iii)) in Section III and the familiar ``pointer states'' that obtain from einselection?}

In the idealized case (e.g., when perfect pointer states exist) the two sets of states are necessarily the same. This is because the key requirement (stability in spite of the monitoring / copying by the environment or an apparatus) that was used in the original definition of pointer states in \cite{69} is essentially identical to ``repeatability'' -- key ingredient of axiom (iii). It follows that when interactions commute with certain observables (e.g., because they depend on them), these observables are constants of motion under such an interaction Hamiltonian, and they will be left intact. For example, interactions that depend on position will favor (einselect) localized states, and destroy (decohere) non-local superpositions. Using {\it predictability sieve} to implement einselection~\cite{5,75,45,52}
is a good way to appreciate this.

{\bf FAQ~\#4}: {\it Repeatability of measurements, axiom (iii), seems to be a very strong assumption. Can it be relaxed (e.g., to include POVM's)?}

Nondemolition measurements are very idealized (and hard to implement). In the interest of brevity we have imposed a literal reading of axiom (iii). This is very much in the spirit of Dirac's textbook, but it is also more restrictive than necessary~\cite{79}, and does not cover situations that arise most often in the context of laboratory measurements. All that is needed in practice is that the {\it record} made in the apparatus (e.g., the position of its pointer) must be ``repeatably accessible''. Frequently, one does not care about repeated measurements of the quantum system (which may be even destroyed in the measurement process). Axiom (iii) captures in fact the whole idea of a record -- it has to persist in spite of being read, copied, etc. So one can impose the requirement of repeatability at the macroscopic level of an apparatus pointer with a much better physical justification than Dirac did for the microscopic measured system. The proof of Section III then goes through essentially as before, but details (and how far can one take the argument) depend on specific settings. This ``transfer of the responsibility for repeatability'' from the quantum system to a (still quantum, but possibly macroscopic) apparatus allows one to incorporate non-orthogonal measurement outcomes (such as POVM's) very naturally: The apparatus entangles with the system, and then acts as an ancilla in the usual projective measurement implementation of POVM's (see e.g.~\cite{NC}).

{\bf FAQ~\#5}: {\it Probabilities -- why do they enter? One may even say that in Everettian setting ``everything happens'', so why are they needed, and what do they refer to?}

Axiom (iii) interpreted in relative state sense ``does the job'' of the collapse part of axiom (iv). That is, when observer makes a measurement of an observable he will record an outcome. Repetition of that measurement will confirm his previous record. That leads to the symmetry breaking derived in Section III and captures the essence of the ``collapse'' in the relative state setting~\cite{75,79}. So, when an observer is about to measure a state (e.g., prepared previously by another measurement) he knows that there are as many possible outcomes as there are eigenvalues of the measured observable, but that he will end up recording just one of them. Thus, even if ``everything happens'', a specific observer would remember a specific sequence of past events that happened to him. The question about the probability of an outcome -- a future event that is about to happen -- is then natural, and it is most naturally posed in this ``just before the measurement'' setting. The concept of probability does not (need not!) concern alternatives that already exist (as in classical discussions of probability, or some ``Many Worlds'' discussions). Rather, (see Ref. \cite{78,Sebens}) it concerns future potential events one of which will become a reality upon a measurement.

{\bf FAQ~\#6}: {\it Derivation of Born's rule here and in~\cite{75,76,78,Z07a}, and even derivation of the orthogonality of outcome states use scalar products. But scalar product appears in Born's rule. Isn't that circular?}

Scalar product is an essential part of {\it mathematics} of quantum theory. Derivation of Born's rule relates probabilities of various outcomes to amplitudes of the corresponding states using symmetries of entanglement. So it provides a connection between mathematics of quantum theory and experiments -- physics. Hilbert space (with the scalar product) is certainly an essential part of the input. And so are entangled states and entangling interactions. They appear whenever information is transferred between systems (e.g., in measurements, but also as a result of decoherence). All derivations proceed in such a way that only two values of the scalar product -- 0 and 1 -- are used as input. Both correspond to certainty. 

{\bf FAQ\#7} {\it How can one infer probability from certainty?}

Symmetry is they key idea. When there are several (say, $n$) mutually exclusive events that are a part of a state invariant under their swaps, their probabilities must be equal. When these events exhaust all the possibilities, probability of any one of them must be $\frac 1 n$. In contrast to the classical case discussed by Laplace, tensor nature of states of composite quantum systems allows one to exhibit {\it objective} symmetries~\cite{75,76,78,Z07a}. Thus, one can dispense with Laplace's  {\it subjective} ignorance (his ``principle of indifference''), and work with objective symmetries of entangled states. The key to the derivation of probabilities are the proofs; (i) That phases of Schmidt coefficients do not matter (this amounts to decoherence, but is established without the reduced density matrix and partial trace, the usual Born's rule - dependent tools of decoherence theory) and; (ii) That equal amplitudes imply equal probabilities. Both proofs~\cite{75,76,78,Z07a} are based on {\it en}tanglement - assisted in{\it variance} (or {\it envariance}). This symmetry allows one to show that certain (Bell state - like) entangled states of the whole imply equal probabilities for local states. This is done using symmetry and certainty as basic ingredients. In particular, one relies on the ability to undo the effect of local transformations (such as a ``swap'') by acting on another part of the composite system, so that the preexisting state of the whole is recovered with certainty. 

One can even use envariance to show that the amplitude of 0 necessarily implies probability of 0 (i.e., impossibility) of the corresponding outcome~\cite{Z07a}. This is because in a Schmidt decomposition that contains $n$ such states with zero coefficients one can always combine two of them to form a new state, which then appears with the other $n-2$ states, still with the amplitude of 0. This purely mathematical step should have no implications for the probabilities of the $n-2$ states that were not involved. Yet, there are now only $n-1$ states with equal coefficients. So the probability $w$ of any state with zero amplitude has to satisfy $nw = (n-1)w$, which holds only for $w=0$. 

One can also prove additivity of probabilities~\cite{78} using modest assumption -- the fact that probabilities of an event and its complement sum up to 1.

{\bf FAQ~\#8}: {\it Why are the probabilities of two local states in a Bell-like entangled state equal? Is the invariance under re-labeling of the states the key to the proof?} 

Envariance is needed precisely because re-labeling is not be enough. For instance, states can have intrinsic properties they ``carry'' with them even when they get re-lableled. Thus, a superposition of a ground and excited states $\ket g + \ket e$ is invariant under re-labeling, but this does not change the fact that the energy of the ground state $\ket g$ is less than the energy of the excited $\ket e$. So there may be intrinsic properties of quantum states (such as energy) that ``trump'' relabeling, and it is {\it a priori} possible that probability is like energy in this respect. This is where envariance saves the day. To see this, consider a Schmidt decomposition of an entagled state 
$\ket \heartsuit \ket \diamondsuit + \ket \spadesuit \ket \clubsuit $ where the first ket belongs to $\cS$ and the second to $\cE$. Probabilities of Schmidt partners must be equal, $p_{\heartsuit} = p_{\diamondsuit}$ and $p_{\spadesuit}=p_{\clubsuit}$. (This ``makes sense'', but can be established rigorously, e.g. by showing that the amplitude of $\ket \clubsuit$ vanishes in the state left after a projective measurement that yields $\heartsuit$ on $\cS$.) Moreover, after a swap $\ket \spadesuit \bra \heartsuit + \ket \heartsuit \bra \spadesuit$, in the resulting state $\ket \spadesuit \ket \diamondsuit + \ket \heartsuit \ket \clubsuit $, one has $p_{\spadesuit} = p_{\diamondsuit}$ and $p_{\heartsuit }=p_{\clubsuit}$. But probabilities in the environment ${\cal E}$ (that was not acted upon by the swap) could not have changed. It therefore follows that $p_{\heartsuit} = p_{\spadesuit}=\frac 1 2$, where the last equality assumes (the usual) normalization of probabilities with $p({\tt certain \ event}) = 1$. 

{\bf FAQ\#9}: {\it Probabilities are often justified by counting, as in the relative frequency approach. Is  counting involved in the envariant approach?}

There is a sense in which envariant approach is based on counting, but one does not count the actual events (as is done is statistics) or members of an imaginary ensemble (as is done in relative frequency approach) but, rather, the number of potential invariantly swappable (and, hence, equiprobable) mutually exclusive events. Relative frequency statistics can be recovered (very much in the spirit of Everett) by considering branches in which certain number of events of interest (e.g., detections of $\ket \heartsuit$, $\ket 1$, ``spin up'', etc.) has occured. This allows one to quantify probabilities in the resulting fragment of the ``multiverse'', with {\it all} of the branches, including the ``maverick'' branches that proved so difficult to handle in the past~\cite{20,21,22,31,48,49,37}. They are still there (as they certainly have every right to be!) but appear with probabilities that are very small, as can be established using envariance~\cite{78}. These branches need not be ``real'' to do the counting -- as before, it is quite natural to ask about probabilities before finding out (measuring) what actually happened.

{\bf FAQ\#10}: {\it What is the ``existential interpretation''? How does it relate to ``Many Worlds Interpretation''?}

Existential interpretation is an attempt to let quantum theory tell us how to interpret it by focusing on how effectively classical states can emerge from within our Universe that is ``quantum to the core''. Decoherence was a major step in solving this problem: It demonstrated that in open quantum systems only certain states (selected with the help of the environment that monitors such systems) are stable. They can persist, and, therefore -- in that very operational and ``down to earth'' sense -- exist. Results of decoherence theory (such as einselection and pointer states) are interpretation independent. But decoherence was not fundamental enough -- it rested on assumptions (e.g., Born's rule) that were unnatural for a theory that aims to provide a fundamental view of the origin of the classical realm starting with unitary quantum dynamics. Moreover, it did not go far enough: Einselection focused on the stability of states in presence of environment, but it did not address the question of what states can survive measurement by the observer, and why. Developments described briefly in this ``guide''  go in both directions. Axiom (iii) that is central in Section III focuses on repeatability (which is another symptom of persistence, and, hence, existence). Events it defines provide a motivation (and a part of the input) for the derivation of Born's rule sketched in Section IV. These two section shore up ``foundations''. Quantum Darwinism explains why states einselected by decoherence are detected by the observers. Thus, it reaffirms the role of einselection by showing (so far, in idealized models) that pointer states are usually reproduced in many copies in the environment, and that observers find out the state of the system indirectly, by intercepting fragments of the environment (which now plays a role of the communication channel). These advances rely on unitary evolutions and Everett's ``relative state'' view of the collapse. However, none of these advances depends on adopting orthodox ``Many Worlds'' point of view, where each of the branches is ``equally real''.

\section{Conclusions}

The advances discussed here include derivation of preferred pointer states (key to postulate (iv)) that does not rely on the usual tools of decoherence,
the envariant derivation of probabilities (postulate (v)), and quantum Darwinism. Taken together, and in the right order, they illuminate the relation
of quantum theory with the classical domain of our experience. They complete the
{\it existential interpretation} based on the operational definition of objective existence, and justify confidence in quantum mechanics as the ultimate theory that needs no modifications to account
for the emergence of the classical.

Of the three advances mentioned above, we have summed up the main idea of the first (the quantum origin
of quantum jumps), provided an illustration of the second (the envariant origin of Born's rule), and briefly
explained quantum Darwinism. As noted earlier, this is not
a review, but a guide to the literature.

Everett's insight -- the realization that relative states settle the problem of collapse -- was the key to these
developments (and to progress in understanding fundamental aspects of decoherence). But it is important to
be careful in specifying what exactly we need from Everett and his followers, and what can
be left behind. There is no doubt that the concept of relative states is crucial. Perhaps even more
important is the idea that one can apply quantum theory to anything -- that there is nothing {\it ab initio}
classical. But the combination of these two ideas does not yet force one to adopt a
``Many Worlds Interpretation'' in which all of the branches are equally real.

Quantum states combine ontic and epistemic attributes. They cannot be ``found out'', so they do not
exist as classical states did. But once they are known, their existence can be confirmed.
This interdependence of existence and information brings to mind two contributions of
John Wheeler:  his early assessment of relative states interpretation (which he saw as
an {\it extension} of Bohr's ideas) \cite{JAW}, and also his ``It from Bit'' program \cite{63} (where information was
the source of existence).

This interdependence of existence and information was very much in evidence in this paper.  Stability
in spite of the (deliberate or accidental) information transfer led to preferred pointer states, and is the
essence of einselection. Entanglement deprives local states of information (which is transferred
to correlations) and forces one to describe these local states in probabilistic terms, leading to Born's
rule. Robust existence emerges (``It from Many Bits'', to paraphrase Wheeler) through quantum Darwinism.  The selective proliferation
of information makes it immune to measurements, and allows einselected states to be
found out indirectly -- without endangering their existence.

I would like to thank C. Jess Riedel and
Michael Zwolak for stimulating discussions. This research was
funded by DoE through LDRD grant at Los Alamos, and, in part, by FQXi.


\end{document}